\documentclass{article}

\usepackage{arxiv}

\usepackage[utf8]{inputenc} 
\usepackage[T1]{fontenc}    
\usepackage{hyperref}       
\usepackage{url}            
\usepackage{booktabs}       
\usepackage{amsfonts}       
\usepackage{nicefrac}       
\usepackage{microtype}      
\usepackage{lipsum}

\usepackage{graphicx}
\usepackage{subcaption}
\usepackage{multirow}
\usepackage{adjustbox}
\usepackage{comment}
\usepackage{authblk}
\usepackage{multicol}
\usepackage{amssymb}
\usepackage{amsmath}
\usepackage{float}

\title{AIforCOVID: predicting the clinical outcomes in  patients with COVID-19 applying AI to  chest-X-rays. An Italian multicentre study.}

\author[1]{Paolo Soda}
\author[2,1]{Natascha Claudia D'Amico}
\author[3]{Jacopo Tessadori}
\author[4,*]{Giovanni Valbusa}
\author[5,1]{Valerio Guarrasi}
\author[6]{Chandra Bortolotto}
\author[3,7]{Muhammad Usman Akbar}
\author[1]{Rosa Sicilia}
\author[1]{Ermanno Cordelli}
\author[2]{Deborah Fazzini}
\author[8]{Michaela Cellina}
\author[8]{Giancarlo Oliva}
\author[6]{Giovanni Callea}
\author[9]{Silvia Panella}
\author[10]{Maurizio Cariati}
\author[11]{Diletta Cozzi}
\author[11]{Vittorio Miele}
\author[12]{Elvira Stellato}
\author[12,13]{Gian Paolo Carrafiello}
\author[14]{Giulia Castorani}
\author[15]{Annalisa Simeone}
\author[6,16]{Lorenzo Preda}
\author[1]{Giulio Iannello}
\author[3]{Alessio Del Bue}
\author[4]{Fabio Tedoldi}
\author[2,4]{Marco Alì}
\author[3,17]{Diego Sona}
\author[2]{Sergio Papa}

\affil[1]{Unit of Computer Systems and Bioinformatics, Department of Engineering, University Campus Bio-Medico of Rome, Via Alvaro del Portillo 21, 00128 Rome, Italy}
\affil[2]{Department of Diagnostic Imaging and Stereotactic Radiosurgey, Centro Diagnostico Italiano S.p.A.,Via S. Saint Bon 20, 20147 Milan, Italy}
\affil[3]{Pattern Analysis and Computer Vision, Istituto Italiano di Tecnologia, Via Morego 30, 16163 Genoa, Italy}
\affil[4]{Bracco Imaging S.p.A., Via Caduti di Marcinelle 13, 20134 Milan, Italy}
\affil[5]{Department of Computer, Control, and Management Engineering, 
Sapienza University of Rome, Via Ariosto, 25, 00185 Rome, Italy}
\affil[6]{Radiology Institute, Fondazione IRCCS Policlinico San Matteo, Viale Golgi 19, 27100 Pavia, Italy}
\affil[7]{Department of Naval, Electrical, Electronic and Telecommunications Engineering - University of Genova, Via All'Opera Pia 11 A, 16145 Genoa, Italy}
\affil[8]{Radiology Department, ASST Fatebenefratelli Sacco,  Piazza Principessa Clotilde 3, 20121 Milan, Italy}
\affil[9]{Diagnostic and interventional radiology unit, ASST Santi Paolo e Carlo - San Paolo Hospital, Via Antonio di Rudinì 8, 20142 Milan, Italy}
\affil[10]{Department of Advanced Diagnostic Technologies - Therapeutic, Diagnostic and Radiology Units, ASST Santi Paolo e Carlo - San Paolo Hospital,Via Antonio di Rudinì 8, 20142  Milan, Italy }
\affil[11]{Department of Emergency Radiology, Careggi University Hospital, Largo Piero Palagi 1, 50139 Florence, Italy}
\affil[12]{Operative Unit of Radiology, Fondazione IRCCS Ca' Granda Ospedale Maggiore Policlinico of Milan, Via della Commenda, 10, 20122 Milan, Italy}
\affil[13]{Department of Health Sciences, Univeristy of Milan, Via Festa del Perdono, 7, 20122 Milan, Italy}
\affil[14]{Diagnostic Imaging, Postgraduate Medical School, University of Foggia,  Via Antonio Gramsci 89, 71122 Foggia, Italy.}
\affil[15]{Department of Diagnostic Imaging, IRCCS Ospedale Casa Sollievo della Sofferenza, Viale Cappuccini 1, 71013 San Giovanni Rotondo, Italy.}
\affil[16]{Radiology Unit, Department of Clinical, Surgical, Diagnostic, and Pediatric Sciences, University of Pavia, Corso Str. Nuova, 65, 27100 Pavia}
\affil[17]{Neuroinformatics Laboratory, Fondazione Bruno Kessler, Via Sommarive, 18, 38123 Trento, Italy}
\affil[*]{Corresponding author: Giovanni Valbusa, email: giovanni.valbusa@bracco.com}

\begin{document}
\maketitle

\begin{abstract}
Recent epidemiological data report that worldwide more than 53 million people have been infected by SARS-CoV-2, resulting in 1.3 million deaths. 
The disease has been spreading very rapidly and few months after the identification of the first infected, shortage of hospital resources quickly became a problem.
In this work we investigate whether chest X-ray (CXR) can be used as a possible  tool for the early identification of patients at risk of severe outcome, like intensive care or death. CXR is a radiological technique that compared to computed tomography (CT) it is simpler, faster, more widespread and it induces lower radiation dose.
We present a dataset including data collected from 820 patients by six Italian hospitals in spring 2020 during the first COVID-19 emergency.
The dataset includes CXR images, several clinical attributes and clinical outcomes.
We  investigate the potential of artificial intelligence to predict the prognosis of such patients, distinguishing between severe and mild cases, thus offering a baseline reference for other researchers and practitioners.
To this goal, we present three approaches that use features extracted from CXR images, either handcrafted or automatically   by convolutional neuronal networks, which are then integrated with the clinical data.
Exhaustive evaluation shows promising performance both in 10-fold and leave-one-centre-out cross-validation, implying that clinical data  and images have the potential to provide useful information for the management of patients and hospital resources.
\end{abstract}


\section{Introduction}\label{sec:Introduction}


According to data reported by the European Centre for Disease Prevention and Control\footnote{ https://www.ecdc.europa.eu/en/geographical-distribution-2019-ncov-cases} as of 13 November 2020 almost 53 million patients worldwide have been infected with the new coronavirus SARS-CoV-2, causing 1.3 million deaths.
Since the identification of patient zero in China, the situation dramatically worsened worldwide, saturating the healthcare system resources. With a shortage of beds available in intensive and sub-intensive care, the need for a quick and effective triage system became an urgency. 

Chest imaging examinations, as chest X-ray (CXR) \cite{bib:schiaffino2020diagnostic} and computed tomography (CT) (\cite{bib:ai2020correlation}) play a pivotal role in different settings. 
Indeed, imaging is used during triage in case of unavailability, delay of or the first negative result of reverse transcriptase-polymerase chain reaction (RT-PCR) \cite{bib:lu2020outbreak}. 
Moreover, imaging is used to stratify disease severity.
Generally, the findings on chest imaging in COVID-19 are not specific and  overlap  with  other  infections. CT should not be used to  screen  for  or as a first-line test to  diagnose  COVID-19  and  should  be  used  sparingly  and reserved for hospitalized, symptomatic patients with specific clinical indications for CT ~\cite{bib:ACR2020}.
CXR most frequent lesions in COVID-19 patients are reticular alteration (up to 5 days from the symptoms onset), and ground-glass opacity (after more than 5 days  from the onset of the symptoms).
In COVID-19 patients' consolidation gradually increase over time. Bilateral, peripheral, middle/lower locations are the most frequent location  \cite{vancheri2020radiographic}. 
In some hospitals, the CXR examination is replaced or accompanied by CT scan, which showed a sensitivity of 97\% for COVID-19 diagnosis \cite{bib:ai2020correlation}, albeit with a limited specificity of 25\%. 
Both CXR and CT have specific pros and cons, but the latter poses several logistic issues, such as the lack of availability of machines’ slots, the difficulty of moving bedridden patients, and the long sanitization times.
Furthermore, patients follow-up with CXR is simplified because it can be acquired at the patient's bed and, when required, directly at home \cite{bib:zanardo2020bringing}.

Recently, artificial intelligence (AI) has been widely adopted to analyse  CXR 
for several purposes, such as tuberculosis detection~\cite{bib:liu2017tx}, abnormality classification and image annotation~\cite{bib:yan2019combining}, pneumonia screening  in pediatric and non pediatric patients~\cite{bib:challenge2018radiological}, edema and fibrosis~\cite{bib:xu2018deepcxray}. 
Obviously, the challenge of COVID-19 pandemic has boosted the research efforts of AI  in medical imaging  and, according to the work presented by Greenspan et.al., \cite{bib:greenspan2020position}, such applications
may have an impact along three main directions, namely, detection and diagnosis, patient management and predictive modelling.

Regarding detection and diagnosis, AI is mainly used to detect the presence of COVID-19 patterns by processing CXR and/or CT images with
deep neural networks (DNNs), such as convolutional neural networks (CNNs). 
DNNs were also applied to lesions segmentation or to produce a coarse localization map of the important regions in the image.
For instance,  Zhang et. al., \cite{bib:zhang2020clinically}  analysed CT scans collected from 4695 patients   to differentiate novel coronavirus pneumonia from other types of pneumonia (bacterial, viral and mycoplasma pneumonia) and from healthy subjects. 
The classification was based on the combination of the segmented lung-lesion map and the normalized CT volumes. 
Experimental tests were performed on 260 patients, achieving an overall accuracy equal to  92.49\%. 
Mineae et.al.\cite{bib:minaee2020deep} analysed 5,000 chest x-ray images from publicly available datasets using four well known convolutional neural networks: ResNet-18, ResNet-50, SqueezeNet, and DenseNet-121. 
Two-thousand images were used for training, whilst the models were tested on the other 3000, attaining a   sensitivity rate equal to  98\%,  and a specificity rate of  around 90\% in  detecting COVID-19 patients from their CXR. 

The  development of systems supporting patient management during the hospitalization is mainly concerned with the monitoring of disease evolution in time.
For instance, Gozes et.al.\cite{bib:gozes2020rapid} proposed an image-based tool supporting the measurement of  disease extent within the lungs.
This severity biomarker is intended to help physicians in the decision-making process by tracking the disease severity over time.

Finally, predictive modeling mainly concerns with the development of models able to predict the progression of the disease.
These approaches usually make use of both imaging  and clinical data to predict the severity of the infection or the progression time, i.e. the time from the initial hospital admission to severe or critical illness, defined by death or the need for mechanical ventilation or the need for being transferred to the intensive care unit (ICU)~\cite{bib:zhang2020clinically}.
Few applications have been recently developed within  this category. 
For example, Greenspan et.al. \cite{bib:greenspan2020position} in their position paper  presented   preliminary and unconsolidated results   on predicting  the probability for a patient to be admitted to the ICU by exploiting quantitative features extracted from the lung region of the CXR images, vital parameters, comorbidities, and other clinical  parameters. 
These data fed a random forest, which attained an  area under the ROC curve (AUC)  equal to 0.83. 
A survey offered by Wynant et.al. \cite{bib:wynants2020prediction},   compared  16 papers presenting  prognostic models  (8 for mortality, 5 for progression to severe/critical state and 3 for length of stay), and the AUC ranged from 0.85  up to 0.99.  
Nine of such papers used only clinical data for the analysis,  whilst the others used clinical data and features extracted from CT images. 
The authors also argued that all   16 papers have a high risk of bias \cite{bib:moons2019probast}   due to the high probability of model overfitting and unclear reporting on intended use of the models.
Still using CT images, two multicentric studies have been recently presented by Yeu et.al.  \cite{bib:yue2020machine} and Chassagnon et.al. \cite{bib:chassagnon2020ai}. 
The former included a cohort of 52 patients from five hospitals to predict short- or long-term hospital stay in patients with COVID-19 pneumonia. 
First, the CT scans were semi-automatically segmented and then for each lesion patch the authors extracted 1,218 features, accounting for first-order, shape, second-order and wavelet measures. 
Second, a logistic model and a  random forest were trained on the data from four hospitals, being tested on patients belonging the fifth clinic. 
They attained balanced accuracies equal to 0.94\% and 0.87\%, respectively.
The work presented by Chassagnon et.al. \cite{bib:chassagnon2020ai} aims to predict patient outcomes (severe vs non-severe) prior to mechanical ventilation support and to suggest a possible prognosis within three available groups (short-term deceased, long-term deceased, long-term recovered).
To these goals they searched for a subset of discriminative features from  several image texture descriptors computed from CT scans and a few clinical data (i.e. age, gender, high blood pressure, diabetes, body mass index). 
On a cohort of 693 patients,  an ensemble of classifiers separated patients with severe vs non-severe outcomes  and it correctly identified the prognosis with    balanced accuracies equal to  70\% and 71\%, respectively. 

\begin{figure}[h]
\centering
\includegraphics[width=1\textwidth]{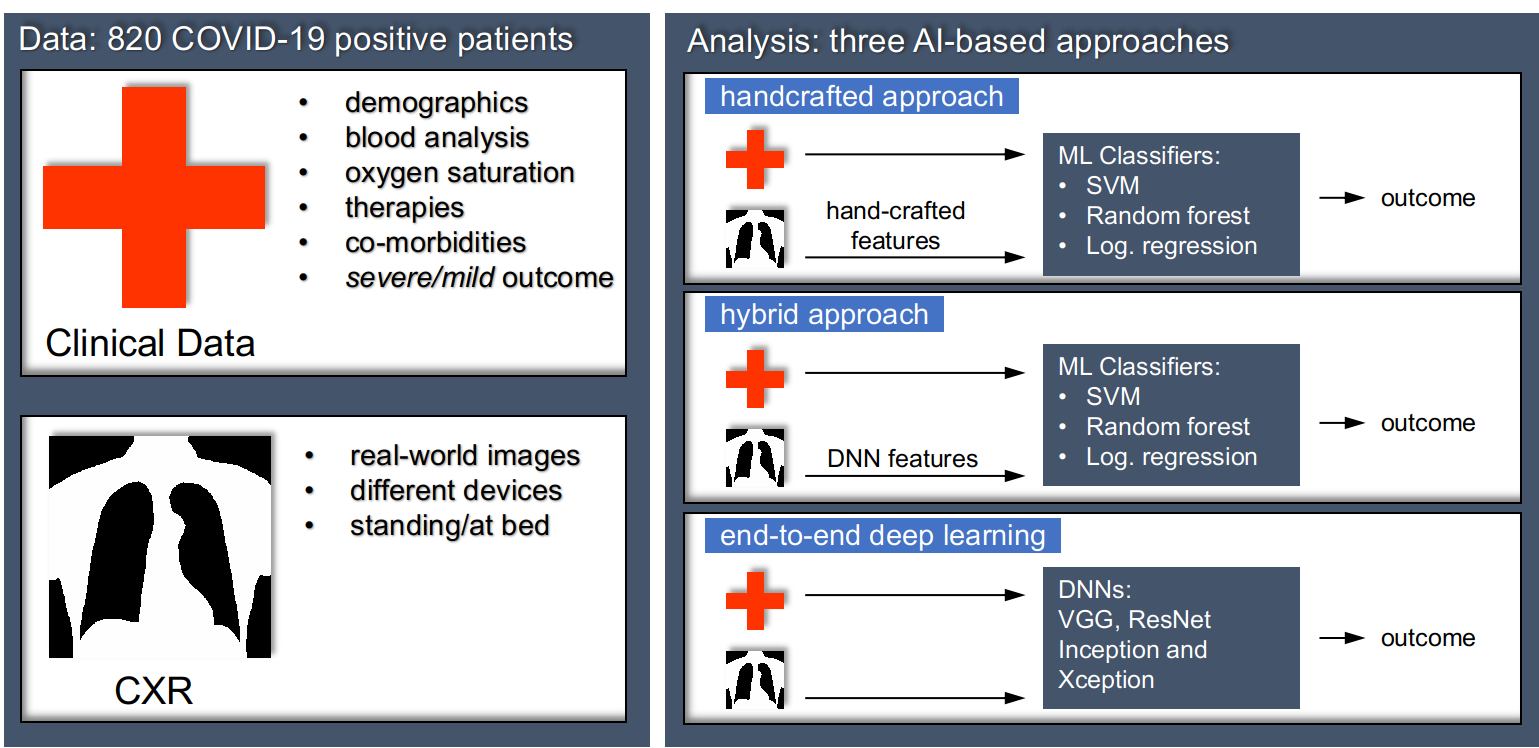}
\caption{Overview of the method for automatic prognosis of COVID-19 in two classes, namely mild and severe. 
Our works includes data collected in 6 independent cohorts, resulting in 820 COVID-19 patients. 
For each, we collected several  clinical  attributes, combined with quantitative  imaging biomarkers computed by handcrafted features or automatically computed by CNNs.
}
\label{fig:Pipeline}
\end{figure}

This analysis of the literature shows that the development of AI-based models predicting the outcomes of COVID-19 patients still deserves further research efforts. 
On the one side,  sharing   patient data from studies as well as creating new data sets collected in clinical practice is fundamental for the AI community \cite{bib:leeuwenberg2020prediction}, since many researchers  do not have the possibility of collecting clinical data and images from different clinical centres.
On the other side, in the  context of COVID-19 prognosis, except for the very preliminary results anticipated by Greenspan et.al. \cite{bib:greenspan2020position}, all the works in literature used CT scans. 
To address both concerns,  this work introduces a novel dataset including clinical data and CXR images from 820 patients with COVID-19 who were hospitalized in six hospitals in Italy.
To each patient we associated prognostic information related to the clinical outcome.
This data repository will be made publicly available to encourage research in this field, where most of resources are collected using CT scans, as already mentioned.
We also investigated three AI-based approaches to predict clinical outcome integrating clinical and imaging data, thus offering a first analysis that can be used by other researchers and practitioners as a baseline reference.
In addition to clinical data, such approaches use quantitative information extracted from the CXR images, which are also referred to as image features or quantitative biomarkers in the following.
The first  approach computes handcrafted texture features to be used by a common classifier, the second approach automatically extracts image descriptors by using a CNN, while the third approach is fully based on DNNs, processing both clinical and image data
(\figurename\ref{fig:Pipeline}).

In synthesis, the main objectives of this work are: 
\begin{enumerate}

    \item to boost the research on AI-based prognostic models to support healthcare systems in the fight against COVID-19 pandemic by making publicly available a repository of CXR images and  clinical data (general information, laboratory data and comorbidities)  collected in a true  environment during the first wave of the pandemic emergency, which include common real-world issues such as missing data, outliers, different imaging devices, poorly standardized data.
    The repository would also facilitate external validation of learning models developed in this field;
    
    \item to present an  evaluation of three  state-of-the-art learning approaches to predict future severe cases at the time of hospitalization,  which are specifically designed to use  either handcrafted or learned image features, together with clinical data. 
    
\end{enumerate}

The rest of this manuscript is organized as follows:  next section describes the dataset we collected and that we are making publicly available. 
Section~\ref{sec:Methods} introduces the methodology we adopted, whilst section~\ref{sec:Result} presents the classification results achieved. 
Section~\ref{sec:discussion} discusses our findings providing  also concluding remarks.

\section{The AIforCOVID dataset} \label{sec:materials}

This study includes the images and clinical data collected in six Italian hospitals at the time of hospitalization of symptomatic patients with COVID-19,  during the first wave of emergency in the country  (March-June 2020) . 
Such data was generated during the clinical activity with the primary purpose of managing COVID-19 patients within the daily practice and they were retrospectively reviewed and collected, after patients' anonymization.
Ethics Committee approval was obtained (Trial-ID: 1507; Approval date: April 7th, 2020) and all data were managed in accordance with the GDPR regulation.
Furthermore, we randomly assigned to each centre a symbolic label, from A up to F.

\begin{table}[t]
\centering
\caption{Patient distribution across the hospitals where the data were collected.}
\label{tab:hospitaldistribution}
\resizebox{.7\textwidth}{!}{
\begin{tabular}{c|c|c}
\toprule
\textbf{Hospital} & \textbf{Number of patients} & \textbf{Mild class prior probability} \\ 
\midrule
A & 120 & 29.2\% \\ \hline
B & 104 & 56.7 \% \\ \hline
C & 31 & 25.8 \% \\ \hline
D & 139 & 54.7 \% \\ \hline
E & 101 & 54.5\% \\ \hline
F & 325 & 46.5\% \\ \bottomrule
\textbf{Total} & \textbf{820} & \textbf{46.8\%} \\ \hline
\end{tabular}}
\end{table}

The 820 CXR examinations reviewed in this study were performed in COVID-19-positive adult patients at the time of  hospital admission (\tablename~\ref{tab:hospitaldistribution}): all  the patients resulted positive for SARS-CoV-2 infection at the RT-PCR test \cite{bib:yang2020clinical}.
In 5\% of such cases the positivity to the swab was obtained only at the second RT-PCR examination. 
In the different centres, CXR examinations were performed using different analog and digital units,
and the execution parameters were settled according to the patient conditions. 
Paired with CXR examinations, we collected also relevant clinical  parameters listed  in~\tablename~\ref{tab:databaseinformation}.

\begin{table}[]
\caption{Description of the clinical data available within the repository.
First and second columns report variables label and description. Summary statistics for the overall population and for the two patients groups are reported in the following columns. 
For continuous variables median and interquartile range are reported, for categorical variables proportions are reported. Feature names followed by '+' were not used for the analysis described in this work.P-values lower than 0.05 were considered significant. * Mann-Whitney U test. $\dagger$ z-test for proportions with Yates continuity correction. $\dagger$ Fisher exact test.  }
\label{tab:databaseinformation}
\begin{adjustbox}{angle=90} 
\resizebox{1.3\textwidth}{!}{%
\begin{tabular}{l|l|l|l|l|l|l}
\toprule
\multirow{2}{*}{\textbf{Name}} & \multirow{2}{*}{\textbf{Description}} & 
\textbf{Overall-} & \multirow{2}{*}{\textbf{Mild-group          (A)}} & \multirow{2}{*}{\textbf{Severe-group     (B)}} & \textbf{A vs B} & 
\textbf{Missing }\\ 
 &  & \textbf{population} &  &  & \textbf{p-value} & \textbf{data (\%)} \\ 
\midrule
\textbf{Active   cancer in the last 5 years} & Patient had active cancer in the last 5 years (\% reported) & 7\% & 5\% & 8\% & \textless{}0.05$\dagger$ & 1.4 \\ \hline
\textbf{Age} & Patient's age (years) & 64; 54-77 & 60; 49-72 & 70; 60-79 & \textless{}0.001* & 0 \\\hline
\textbf{Atrial Fibrillation} & Patient had atrial fibrillaton (\% reported) & 9\% & 5\% & 11\% & \textless{}0.01 $\dagger$ & 2.2 \\ \hline
\textbf{Body temperature ($^{\circ}$C)} & Patients temperature at admission (in $^{\circ}$C) & 38; 37-38 & 38; 37-38 & 38; 37-38 & 0.171 & 8.8 \\ \hline
\textbf{Cardiovascular Disease} & Patient had cardiovascular diseases (\% reported) & 35\% & 23\% & 40\% & \textless{}0.001$\dagger$ & 1.7 \\ \hline
\textbf{Chronic Kidney disease} & Patient had chronic kidney disease (\% reported) & 6\% & 4\% & 9\% & \textless{}0.01$\dagger$ & 1.4 \\ \hline
\textbf{COPD} & Chronic obstructive pulmonary disease (\% reported) & 7\% & 4\% & 10\% & \textless{}0.01$\dagger$ & 1.4 \\ \hline
\textbf{Cough} & Cought (\%yes) & 54\% & 59\% & 50\% & \textless{}0.05$\dagger$ & 0.5 \\ \hline
\textbf{CRP} & C-reactive protein concentration (mg/dL) & 57; 24-119 & 42; 17-75 & 103; 48-163 & \textless{}0.001* & 3.5 \\ \hline
\textbf{Days Fever} & Days of fever up to admission (days) & 3; 2-4 & 3; 2-4 & 3; 2-3 & 0.289 & 10.96 \\  \hline
\textbf{D-dimer} & D-dimer amount in blood & 632; 352-1287 & 549; 262-909 & 820; 438-2056 & \textless{}0.001* & 77.6 \\ \hline
\textbf{Death+} & Death of patient occurred during hospitalization for any cause & 168 & 0 & 168 & - & - \\ \hline
\textbf{Dementia} & Patient had dementia (\% reported) & 4\% & 3\% & 6\% & 0.087 & 1.8 \\ \hline
\textbf{Diabetes} & Patient had diabetes (\% reported) & 16\% & 10\% & 21\% & \textless{}0.001$\dagger$ & 1.4 \\ \hline
\textbf{Dyspnea} & Patient had intense tightening in the chest, air hunger, difficulty & 50\% & 37\% & 62\% & \textless{}0.001$\dagger$ & 0.4 \\
&  breathing, breathlessness or a feeling of suffocation (\%yes)& & & & & \\ \hline
\textbf{Fibrinogen} & Fibrinogen concentration in blood (mg/dL) & 607; 513-700 & 550; 473-658 & 615; 549-700 & \textless{}0.001* & 73.6 \\ \hline
\textbf{Glucose} & Glucose concentration in blood (mg/dL) & 110; 96-130 & 104; 93-121 & 114; 101-139 & \textless{}0.001* & 20.6 \\ \hline
\textbf{Heart Failure} & Patient had heart failure (\% reported) & 2\% & 1\% & 3\% & 0.157 & 2.3 \\ \hline
\textbf{Hypertension} & Patient had high blood pressure (\% reported) & 46\% & 38\% & 54\% & \textless{}0.001$\dagger$ & 1.4 \\ \hline
\textbf{INR} & International Normalized Ratio & 1.13; 1.07-1.25 & 1.11; 1.06-1.20 & 1.15; 1.08-1.28 & 0.004* & 28.8 \\\hline 
\textbf{Ischemic Heart Disease} & Patient had ischemic heart disease (\% reported) & 15\% & 11\% & 18\% & \textless{}0.01$\dagger$ & 18.3 \\ \hline
\textbf{LDH} & Lactate dehydrogenase concentration in blood (U/L) & 320; 249-431 & 271; 214-323 & 405; 310-527 & \textless{}0.001* & 24.6 \\ \hline
$\mathbf{O_2}$ \textbf{(\%)} & Oxygen percentage in blood (\%) & 95; 90-97 & 96; 94-98 & 92; 87-96 & \textless{}0.001* & 16.5 \\ \hline
\textbf{Obesity} & Patient had obesity (\% reported) & 9\% & 6\% & 11\% & 0.058 & 36.1 \\ \hline
$\mathbf{PaCO_2}$ & Partial pressure of carbon dioxide in arterial blood (mmHg) & 33; 30-36 & 34; 30-37 & 33; 30-35 & 0.116 & 15.4 \\ \hline
$\mathbf{PaO_2}$ & Partial pressure of oxygen in arterial blood (mmHg) & 69; 59-80 & 73; 67-81 & 64; 54-76 & \textless{}0.001* & 15.3 \\ \hline
\textbf{PCT} & Platelet count (ng/mL) & 0.19; 0.09-0.56 & 0.09; 0.05-0.26 & 0.28; 0.13-0.72 & \textless{}0.001* & 71.8 \\ \hline
\textbf{pH} & Blood pH & 7; 7-7 & 7; 7-7 & 7; 7-7 & \textless{}0.001* & 17.3 \\ \hline
\textbf{Position+} & Patient position during chest x-ray (\%supine) & 78\% & 68\% & 87\% & \textless{}0.001$\dagger$ & 0 \\\hline
\textbf{Positivity at admission} & Positivity to the SARS-CoV-2 swab at the admission time  & 95\% & 94\% & 96\% & 0.142 & 4.7 \\ 
& (\%   positive)& & & & &\\ \hline
\textbf{Prognosis} & Patient outcome, see section \ref{sec:materials}  (\% cases)& - & 46.8\% & 53.2\% & 0.468$\dagger$ & 0.0 \\
\hline
\textbf{RBC} & Red blood cells count (10\textasciicircum{}9/L) & 4.65; 4.26-5.07 & 4.70; 4.34-5.11 & 4.59; 4.13-5.03 & \textless{}0.001* & 3.0 \\ \hline
\textbf{Respiratory Failure} & Patient had respiratory failure (\% reported) & 1\% & 100\% & 2\% & 0.131 & 19.0 \\ \hline
$\mathbf{SaO_2}$ & arterial oxygen saturation (\%) & 95; 91-97 & 96; 94-98 & 92;87-96 & \textless{}0.001* & 59.2 \\ \hline
\textbf{Sex} & Patient's sex (\%males) & 68\% & 59\% & 75\% & \textless{}0.001$\dagger$ & 0 \\ \hline
\textbf{Stroke} & Patient had stroke (\% reported) & 4\% & 3\% & 4\% & 0.447 & 2.3 \\ \hline
\textbf{Therapy Anakinra+} & Patient was treated with Anakinra (\%yes) & 100\% & 0\% & 0\% & - & 10.8 \\ \hline
\textbf{Therapy anti-inflammatory+} & Patient was treated with anti-inflammatory drugs & 55\% & 53\% & 57\% & 0.243 & 13.5 \\
&therapy   (\%yes) & & & & & \\ \hline
\textbf{Therapy antiviral+} & Patient was treated with antiviral drugs (\%yes) & 47\% & 44\% & 50\% & 0.129 & 10.7 \\ \hline
\textbf{Therapy Eparine+} & Patient was treated with eparine   (no; yes; prophylactic  & 56.6\%;	11.5\%;	 &	73.3\%;	8.3\%; &	39.9\%;	14.7\%;	
 & \textless{}0.001$\dagger$& 13.4 \\ 
 & treatment; therapeutic treatment) &28\%;	3.9\% & 	17.2\%;	1.1\%&38.8\%;	6.6\% & & \\ \hline
\textbf{Therapy hydroxychloroquine+} & Patient was treated with hydroxychloroquine (\%yes) & 59\% & 56\% & 62\% & 0.118 & 11.6 \\\hline
\textbf{Therapy Tocilizumab+} & Patient was treated with Tocilizumab (\%yes) & 9\% & 2\% & 15\% & \textless{}0.001$\dagger$ & 12.4 \\ \hline
\textbf{WBC} & White blood cells count (10\textasciicircum{}9/L) & 6.30; 4.73-8.42 & 5.58; 4.32-7.17 & 7.10; 5.25-9.80 & 0.012* & 0.7\\
\bottomrule
\end{tabular}}
\end{adjustbox}
\end{table}

\begin{figure}[h]
\centering
\includegraphics[width=0.7\textwidth]{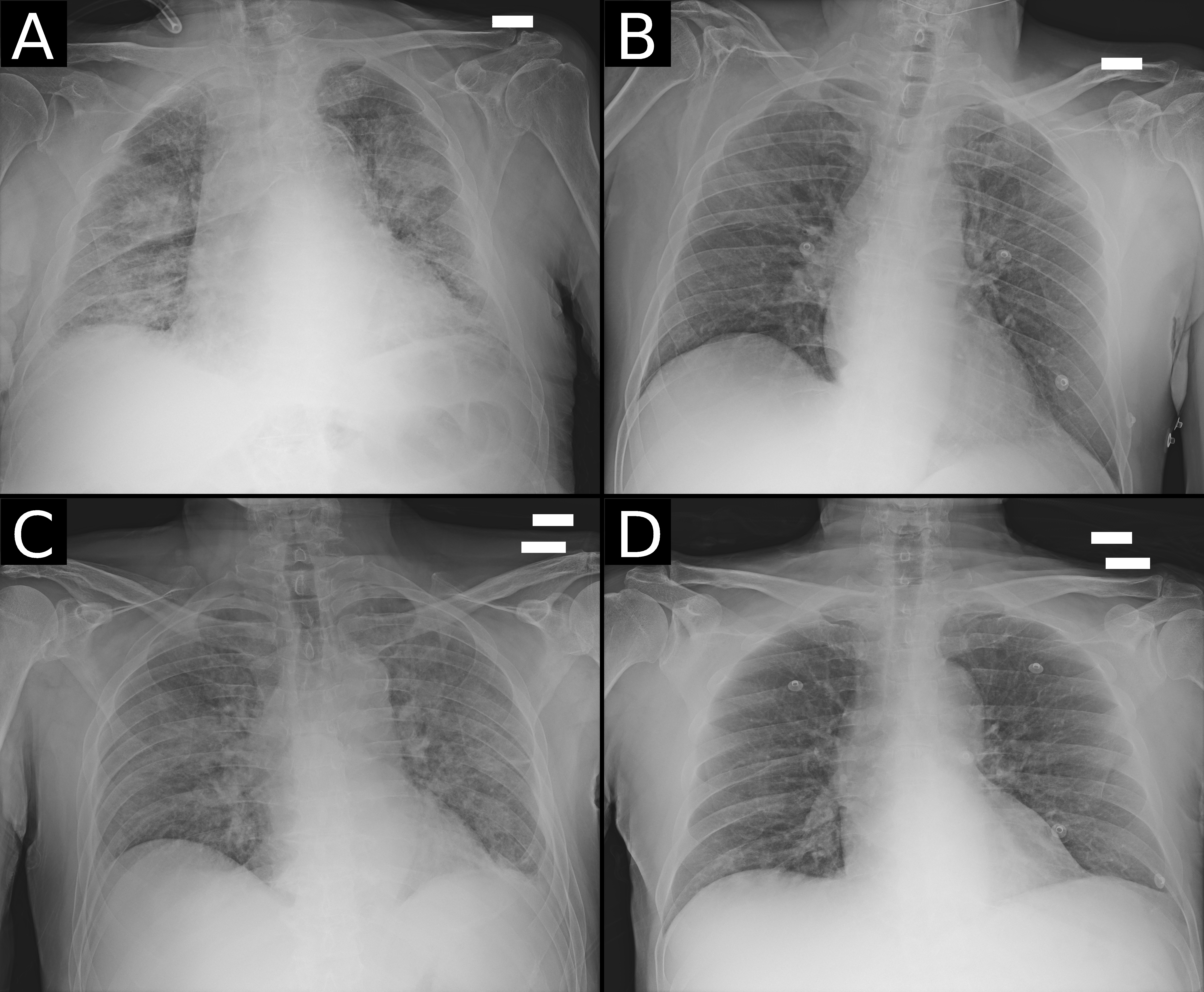}
\caption{Examples of CRX images of  patients with COVID-19 available within the dataset. 
Panels A and B show two images of patients with severe outcome whilst the radiological visual inspection may suggest severe and mild prognoses, respectively. 
Similarly, panels C and D show two images of patients with mild outcome whilst a radiologist may suggest severe and mild prognosis, respectively, on the basis of the visual interpretation.}
\label{fig:exampleRX}
\end{figure}

According to the clinical outcome, each patient was assigned to the {\em mild} or the {\em severe} group. 
The former contains the patients sent back to domiciliary isolation or hospitalized without ventilatory support, whereas the latter is composed of patients who required non-invasive ventilation support, intensive care unit (ICU) and deceased  patients. 
\figurename~\ref{fig:exampleRX} shows four difficult examples of CXR images within the  dataset: indeed, panels A and B show two images of patients with severe outcome whilst the radiological visual inspection may suggest severe and mild prognoses, respectively. 
Similarly, panels C and D show two images of patients with mild outcome whilst a radiologist may report severe and mild prognosis, respectively.

During an initial data quality cleaning, we double-checked  with the clinical partners the anomalous data and the outliers, i.e. those values lying outside the expected clinical range  or identified applying the interquartile range method, which were then corrected when needed. 
Categorical variables  values were homogenized to a coherent coding, such as 0 and 1 values for binary variables like comorbidities and sex, and we adopted the string ``NaN'' to denote missing data. 
No exclusion rule was applied for images based on device type or brand (e.g. digital or analog devices) or patient positions (standing or at bed), whereas X-ray images taken with lateral projection were excluded. 
In the case of multiple CXR images delivered for the same patient, 
the dataset contains only the first one. 
It is worth noting that the presence of   missing entries in the clinical data mostly depends upon  the procedures carried out in the individual hospitals as well as  upon the pressure due to the overwhelming number of patients hospitalized during the COVID-19 emergency. 
For the sake of completeness, the rate of missing data is reported in the last column of~\tablename~\ref{tab:databaseinformation}.

CXR images   were    collected in DICOM format and, for anonymization constrains, all the fields but a set of selected metadata related to acquisition parameters were blanked in the DICOM header (e.g. image modality, allocated  bits, pixel spacing, etc.). 
All the images in the repository are currently stored using 16 bits, while acquisition precision varies: 13.5\% were acquired at 10 bits precision, 35.4\% at 12 bits, 46.6\% at 14 bits and 4.5\% using the full 16 bits precision.
Furthermore, all the images were acquired with isotropic pixel spacing ranging from 0.1 mm to 0.2 mm. 
The most common pixel spacing is 0.15 mm, 0.1 mm and 0.16 mm for 43.9\%, 13.7\% and 13.6\% of images respectively. 
Image sizes, in pixels, are distributed as follows: 33.4\% of the images have 2336$\times$2836 pixels, 13.5\% of images have 3520$\times$4280 pixels and 10.1\% of the images have 3480$\times$4240 pixels. 
The other images have a number of rows  ranging from 1396 up to 4280, whilst the number of columns ranges from 1676 up to 4280.

\subsection{Statistical analysis of clinical data}
\label{sec:StatisticalAnalysis}

We performed a statistical analysis applying the Mann-Whitney U test to compare mild- and severe-groups in case of continuous variables, whereas we used the z test with Yates continuity correction for analysing proportions. 

Summary statistics are reported in~\tablename~\ref{tab:databaseinformation}. 
For continuous variables median and interquartile range (IQR) were reported. 
For categorical variables we reported patients' proportions expressed as percentage. 
For statistical analysis a p-value lower than 0.05 was considered significant.

\begin{figure}[h]
\centering
\resizebox{0.99\columnwidth}{!}{
\includegraphics{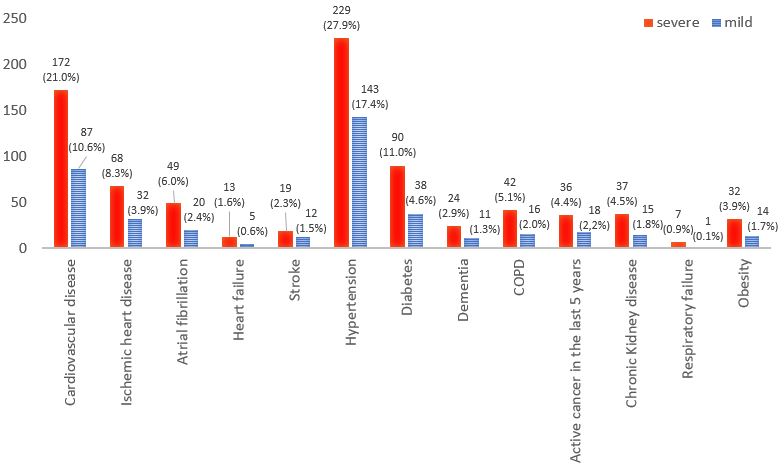}}
\caption{Comorbidity distributions between groups. For all data, value and percentage referred to the total population was indicated.}
\label{fig:diseasedistribution}
\end{figure}

The  analysis evidenced  that females represented the  32\%  (n=266) of the total population and they were significantly ($p <$0.001) older (median age 70 years, IQR 57–80 years) than males (median age 64 years, IQR 53–74 years).
Furthermore, 522 out of 820 (63\%) patients had at least one comorbidity (~\figurename~\ref{fig:diseasedistribution}). 

In agreement with widely reported demographic data showing that older patients had more severe outcome, in our dataset the patients of severe-group (70 years, IQR 60–79) were significantly ($p<$0.001) older than those belonging to the mild-group (60 years, IQR 49–72 years). 
Three hundred twenty patients out of 436 (73\%) of the severe group had at least one comorbidity; in the mild-group they were 194 out of 384 (51\%).

Moreover, 47\% (384/820) of patients belonged to the mild-group, of which 157 (41\%) were females and significantly ($p=0.015$) older (median 63 years, IQR 50–76 years) than males (median 59 years, IQR 48–69 years). 
In the severe-group were 436 out of 820 (53\%) patients, of which 109 (25\%) were females and significantly ($p<$0.001) older (median age 78 years, IQR 67–85 years) than males (median age 67 years, IQR 57–76 years). 
Severe group consisted of 43\% (189/436) of patients hospitalized with non-invasive ventilation support, 18\% (79/436) of patients in ICU, and 38\% (168/436) of dead patients.
Regarding the dead patients' subgroup, the mean age was 78 (IQR 68–84) years: the youngest was 43yo while the oldest 97yo. 
Among dead patients, 97\% of them had at least one co-morbidity, while 17\% had five comorbidities reported. 
The majority (72\%, 121 out 168) of dead patients were male.

\section{Methods} \label{sec:Methods}

We investigated  three  AI-based prognostic approaches covering  well-known methodologies with the intent  to offer to researchers and practitioners a reference baseline to process the data available within the  AIforCovid dataset.
 Furthermore, for the sake of an easy and fair comparison and to foster further research in this field, we detail also the adopted validation procedures, recommending others to measure models performance at least as reported here.

As schematically depicted in~\figurename~\ref{fig:Pipeline}, the first learning approach employs first order and  texture features computed from the images, which are mined together with the clinical data feeding a supervised learner.
In the following, it is shortly referred to as {\em handcrafted approach}, and it is presented in section~\ref{subsec:handcrafted}.

In the last decade we have assisted to the rise of  deep artificial neural networks, which have attained outstanding performance in many fields.
Recently, DNNs such as convolutional neural networks have been applied also to COVID-19 imaging mostly for diagnostic purposes \cite{bib:greenspan2020position}.
On this basis, the second approach presented here mixes together automatic features computed by a CNN with the clinical data. 
Shortly, we used a pre-trained CNN as a CXR feature extractor. The output of the last fully-connected layer was then provided as input for a SVM classifier, together with the clinical features.

In the following, it is shortly referred to as {\em hybrid approach}, and it is presented in section~\ref{subsec:hybrid}.

The third approach exploits together  the  clinical data and the raw CXR using a multi-input convolutional network to predict patients’ outcome. 
In order to handle data from such different sources, the network consists of two dedicated input branches, while higher-level features from both sources are concatenated in the last layers before the classification output. 
 In the following, this approach is shortly referred to as {\em end-to-end deep learning  approach}, and it is detailed  in section~\ref{subsec:endDL}.

Note that all such approaches do not use the therapy-related variables included in the dataset because, albeit therapy could influence the final outcome, it is also dependent on the outcome (i.e. patients which required intensive care were administered with specific therapies). 
Furthermore, the  classification task defined considers only the data collected at the time of hospitalization and, therefore, in a true clinical scenario, information on the administered therapy would not be available.
For this reason, the use of those variables could be misleading.

Before presenting in detail each of the three approaches,  following sections~\ref{subsec:dataImputation} and~\ref{subsec:imageStandardization} describe   data imputation and image standardization. 
Furthermore, section~\ref{subsec:LungSegmentation} presents the  framework used to segment the lung, whereas  section~\ref{subsec:FeatureSelectionClassifiers} describes the feature selection approach and the classifiers adopted, which are the same across the three methods to  facilitate their comparison. 
\tablename~\ref{tab:SintesiProcessing} summarize the common operations applied by each of the three AI methods and  section~\ref{subsec:modelValidation} introduces the procedure adopted to validate the learning models.

\begin{table}[th]
\centering
\caption{Summary of the operations common to the three AI approaches.}
\label{tab:SintesiProcessing}
\resizebox{.99\textwidth}{!}{%
\begin{tabular}{lcccc}
\toprule
\multirow{2}{*}{\textbf{Method}} & \multicolumn{4}{c}{\textbf{Operations}}                                                                              \\ \cline{2-5}
                                 & \textit{Data imputation}  & \textit{Image standardization} & \textit{Lung segmentation} & \textit{Feature selection} \\  \midrule
Handcrafted & \checkmark & \checkmark & \checkmark & \checkmark \\
Hybrid  & \checkmark & \checkmark      & \checkmark  & \checkmark  \\
End-to-end DL & \checkmark & \checkmark      &                            &     \\
\bottomrule

\end{tabular}}
\end{table}

\subsection{Data imputation} \label{subsec:dataImputation}

To deal with missing data, univariate data imputation  estimates missing entries by using the  mean of each column in which the missing values are located.
We preferred this approach to  multivariate or prediction-based imputation methods since it is known to work well when the data size is not very large, 
and it can prevent data loss which results from brute force  rows and columns removal. 
Furthermore, preliminary results not shown here confirmed such observations.
As reported in the second column of~\tablename~\ref{tab:SintesiProcessing},  imputation was performed before each   learning paradigm worked on the data.

\subsection{Images standardization}\label{subsec:imageStandardization}

CXR images collected for this study were acquired with different devices and acquisition conditions, 
as mentioned in section~\ref{sec:materials}. 
For this reason, we applied image normalization that, to a large extent, is the same for all the  three methods.
Indeed, for the handcrafted approach pixels values were normalized to have zero mean and unit  standard deviation, whilst the images were resized to 1024$\times$1024 pixels using bilinear interpolation. 
For the hybrid approach, a segmentation network was used to identify the square box containing the lungs, in a way to crop only the region of interest, as detailed in the next section. 
The images were then normalized and resized to a dimension equal to 224$\times$224 pixels,  as we  employed early processing layers pre-trained on the ImageNet dataset, which consists of images of this size. 
Similarly, in the end-to-end DL approach images were resized to 224$\times$224 pixels, without prior cropping, and normalized as in the previous cases. 

\subsection{Lung segmentation}\label{subsec:LungSegmentation}
\begin{figure}[h]
  \centering
  \includegraphics[width=0.4\textwidth]{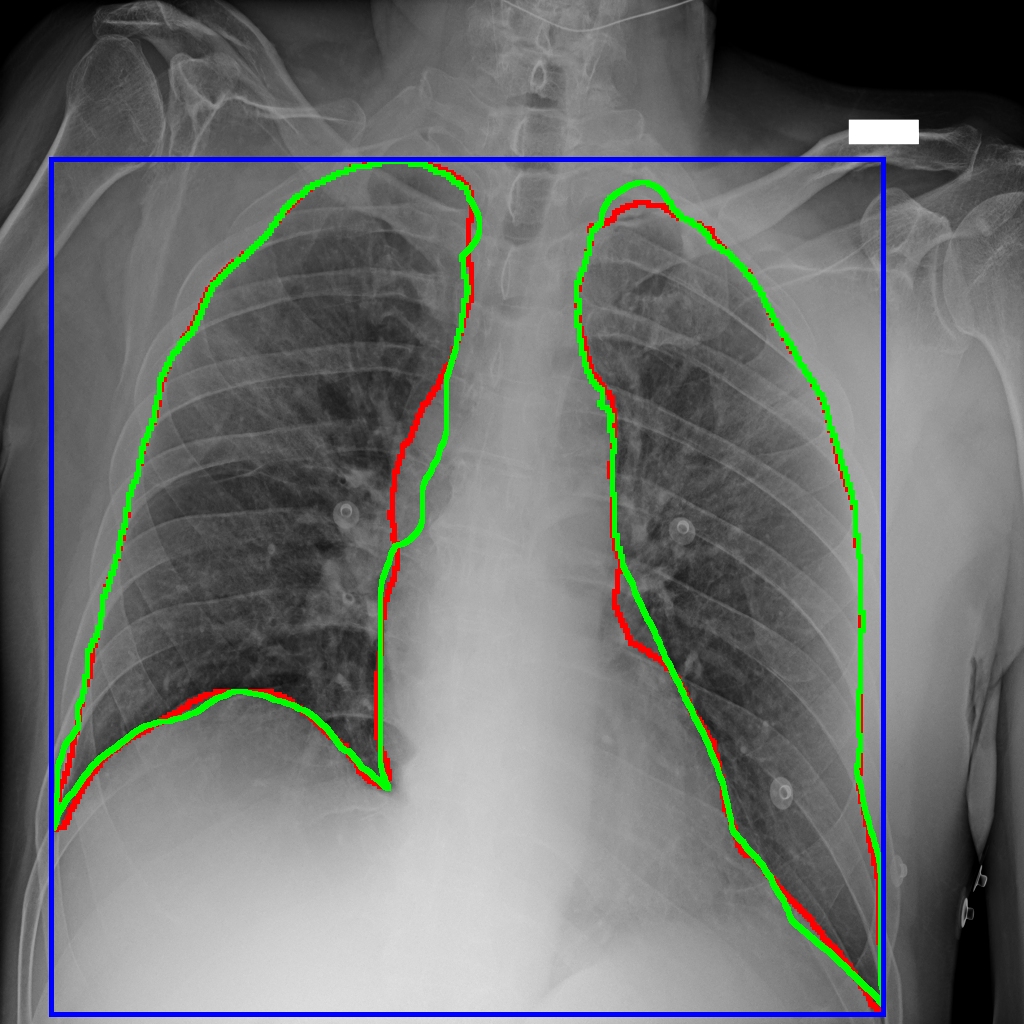}
  \caption{Example of the lung segmentation results. Green line: manual segmentation, red line:  segmentation returned  by  the U-Net, blue line:  bounding box from U-Net  segmentation. }
  \label{fig:LungSegmentation}
\end{figure}

When needed, to segment the lung we apply a semi-automatic approach that initially delineates the lung borders using a U-Net,  which is a convolutional neural network architecture for fast and precise segmentation of images. 
The network was already trained  on non-COVID-19 lung CXR datasets\footnote{The network is available as detailed in the reference denoted as \cite{bib:lungseg}.}, namely the Montgomery County  CXR  set (MC) presented by Jaeger et.al. \cite{bib:jaeger2014two}  and the Japanese Society of Radiological Technology (JSRT) repository  presented by Shiraishi et.al. \cite{bib:shiraishi2000development}. 
The former contains 7,470 CXR  collected by the National Library of Medicine within the Open-i service, whereas the latter is composed of 247 chest radiographs with and without a lung nodule. 
The U-net requires input images represented as 3-channel 256$\times$256 matrices and, hence,  grayscale images were copied to all the channels and then resized.
Furthermore, we normalized the pixel intensities as detailed in section~\ref{subsec:imageStandardization}. 
After these transformations, each image was passed through the convolutional network and all the pixels were classified as foreground (i.e. the lung) or as background. 
To check if the network worked well, hand-made masks were segmented by expert radiologists (\figurename~\ref{fig:LungSegmentation}). 

\subsection{Feature selection and classifiers} \label{subsec:FeatureSelectionClassifiers}

In general, we had a  large number of descriptors that suggested us to apply a feature selection stage, which was set up in  two steps.
The first is a coarse step that runs a univariate  filtering based on mutual information as score function to pre-select a reduced set of image descriptors, whatever the approach used for their computation. 
The calculation of mutual information between continuous features with the discrete class variable was addressed estimating the entropy from the k-nearest neighbours distances \cite{bib:mutualinfocontinuousvar}. 

The second feature selection step merges the pre-selected imaging features with the clinical data. 
To this end, we applied a wrapper approach, namely the  Recursive Feature Elimination and Cross-Validated selection (RFECV) method \cite{bib:guyon2002gene}, which receives as input  the  pre-selected imaging descriptors and   the 34 clinical features.
Indeed, the RFECV is fed by an increasing number of pre-selected imaging descriptors ($D_{pr}$): 
fine-grained sampling was carried out for $D_{pr}$ $\leq$ 10 applying a step of 2; for 10 $<$ $D_{pr}$ $\leq$ 50, $D_{pr}$ was sampled with step of 5; 
finally, RFECV was fed with all the image features. 
RFECV applies a pruning procedure that starts considering all features in the dataset and recursively eliminates the less important according to a feature's importance score calculated using a classifier. 
Note that the optimal number of features is selected by RFECV using nested  5-fold cross-validation on the training set.

With reference to the base learner we evaluated three different computational paradigms: Logistic Regression (LGR); Support Vector Machines  with a linear kernel (SVM); and Random Forests (RF).
For all parameters in the adopted models we used the default values provided by the libraries, without any fine tuning. Indeed, we were not interested on the best absolute performance. Moreover,  Arcuri et.al.
\cite{arcuri2013parameter} empirically observed that in many cases the use of tuned parameters cannot significantly outperform the default values of a classifier suggested in the literature.

\subsection{Models validation}\label{subsec:modelValidation}
Model validation for the three tested methods consists of k-fold and leave-one-centre-out cross validation.
For each cross-validation run, the training fold was used for data normalization, parameters' estimation and/or features ' selection depending on the applied method.
Classification performance assessment was carried out using testing fold data only; k-fold cross-validation was repeated with k equal to 3 and 10 with 20  repetitions.   
In leave-one-centre-out (LOCO) cross validation, in each run the test set is composed of all the samples belonging to one centre only, while the others were assigned to the training set. 
When needed, the validation set was extracted from the training set using any  policies (such as random selection, hold-out, nested cross validation, etc.), and  considering also the computational burden.

Performance of the learning models were measured in terms of accuracy, sensitivity and specificity, reporting the average and standard deviation of each experiments. 
When needed, we ran the pairwise two-sided Mann Whitney U test to compare  the results provided by two methods,  whereas we performed the Kruskal-Wallis test followed by the Dunn's test with Bonferroni correction for multiple comparisons. In the rest of the manuscript we assume that the pairwise two-sided Mann Whitney U test was  performed by default, otherwise we will specify the test used.  

\subsection{Handcrafted approach} \label{subsec:handcrafted}
The handcrafted approach first computes  parametric maps of the  lungs segmented in the CXR image;  second it extracts several features that are then provided together with the clinical data to a supervised learner.

To segment the lung we applied the approach presented in section~\ref{subsec:LungSegmentation},  comparing also the segmentation masks provided by the U-Net with those manually annotated. 
We found that  the network provides a Jaccard index and a Dice score equal to 0.896 and  0.942, respectively.
We deem that such  performance are satisfactory as long since it is only needed to recover the bounding box, as in the hybrid approach presented below, while it would not be sufficient for exact lung delineation.  
For this reason, the lung masks obtained so far are then reviewed by an expert radiologist  and then used to compute the handcrafted features as follows.

From the segmented lungs we computed the parametric maps using a pixel-based approach as proposed by  Penny et.al. \cite{bib:penny2011statistical}.
Pixels values of the parametric maps were obtained by computing first- and second-order radiomic features on a 21x21 sliding window running over each pixel of the entire lung region.
First-order measures   describe the statistical distribution of tissue density inside the kernel; from its grey levels’ histogram, we extracted 18 descriptors, 
whose formal presentation is offered in  \ref{appendix}. 
Second-order descriptors are based on the Grey Level Co-occurence Matrix (GLCM): at each location, we got a GLCM image, where we computed 24 Haralick descriptors \cite{bib:haralick1973textural}   
detailed in the same Appendix as before.
This procedure returned  42 parametric images (18 First-order + 24 GLCM) 
for each CXR image, where we finally computed seven  statistics, namely: mean, median, variance, skewness, kurtosis, energy and entropy.
This resulted in  294 image features (i.e. 7 statistics by 42 parametric maps). 

To cope with the  large number of descriptors we proceeded as described in section \ref{subsec:FeatureSelectionClassifiers}, adopting the base learners already described there. 
Then, for each  tested classifier, given the set and number of descriptors selected by the wrapper approach in the nested cross-validation fashion, we trained the same classifier on the whole training fold and measured  recognition performance on the test fold.

\subsection{Hybrid approach} \label{subsec:hybrid}
The hybrid approach integrated the output of a pre-trained deep network and the set of clinical measures. 
The pipeline worked as follows: first, we applied a pre-trained deep neural network to segment the lungs; second, a convolutional neural network was trained to extract relevant features from the CXR images; third, we concatenated the deep features with the clinical ones; fourth, we performed a feature selection step 
as reported in section~\ref{subsec:FeatureSelectionClassifiers}; fifth, we trained a supervised classifier to accomplish the binary classification task. 
In the following we will illustrate these steps. 

As mentioned before, the image repository is composed of CXR images collected in multiple hospitals, using  different machines with different acquisition parameters. 
This resulted in a certain degree of variability among the images, where the lungs have also different sizes.
To cope with this issue, we adopted the segmentation net already discussed in section~\ref{subsec:LungSegmentation},  which boosts the performance of the feature extraction network by locating the lungs.
Differently from before, where the U-Net was used to pre-segment the lungs whose borders were manually refined, here we adopted a fully automated approach since the segmentation mask given by network was used to extract the rectangular bounding box containing the ROI. 
Now there was no need for any manual intervention since the  performance  at the level of ROI bounding box segmentation was satisfactory, when compared with human's annotation. 
Indeed, the Jaccard index and the Dice score were now equal to 0.929 and 0.960, respectively\footnote{In only one case the segmentation network did not segment the lungs; in this case, the entire original image is used.}.

Next, each ROI was resized to a square so that the longest side of the ROI was mapped to the square side, and the other ROI side was resized accordingly. 
Each cropped image was then passed to a deep neural network to extract the features, where we performed  a  transfer learning process as follows. 
Indeed, preliminary experiments showed that such an approach gave better results than starting the training from scratch.
On our image dataset we trained  several state-of-the-art network architectures previously initialized on other repositories. 
In a first stage we tested in ten-fold cross validation these networks: Alexnet \cite{bib:krizhevsky2014one}, VGG-11, VGG-11 BN, VGG-13, VGG-13 BN, VGG-16, VGG-16 BN, VGG-19, VGG-19 BN \cite{bib:simonyan2014very}, ResNet-18, ResNet-34, ResNet-50, ResNet-101, ResNet-152 \cite{bib:he2016deep}, ResNext \cite{bib:xie2017aggregated}, Wide ResNet-50 v2 \cite{bib:zagoruyko2016wide}, SqueezeNet-1.0, SqueezeNet-1.1 \cite{bib:iandola2016squeezenet}, DenseNet-121, DenseNet-169, DenseNet-161, DenseNet-201 \cite{bib:huang2017densely}, GoogleNet \cite{bib:szegedy2015going}, ShuffleNet v2 \cite{bib:ma2018shufflenet} and MobileNet v2 \cite{bib:sandler2018mobilenetv2}. 
Then, to reduce the computational burden, the top-five networks (i.e. VGG-11, VGG-19, ResNet-18, Wide ResNet-50 v2, and GoogleNet)  underwent all the experiments described in section \ref{subsec:modelValidation}.
In all the cases, we changed the output layer of the CNNs, using two neurons, one for each class. 
Moreover, image standardization as described in section~\ref{subsec:imageStandardization} was performed.  
We also augmented the training  data by independently applying the following transformations    with a probability equal to 30\%: vertical and horizontal shift (-7, +7), y-axis flip, rotation (-175$^\circ$, +175$^\circ$) and elastic deformation ($\sigma$ =7, $\alpha$ = [20,40]). 
Training parameters were:  a batch size of 32  with a cross-entropy loss, a SGD optimizer with learning rate of 0.001 and momentum of 0.9, with  max epochs sets equal to  300 and an early stopping criterion fixed at 25 epochs, using the accuracy on the validation set.

Once the deep networks were trained, we integrated the automatic features they computed with the clinical information. 
To this goal, we extracted the last fully connected layer for each network, which was used as a vector of features for each patient; accordingly, on the basis of the network we were using, the number of automatic features varied between 512 and 4096 (i.e. it is 512 for ResNet-18, 1024 for GoogleNet, 2048 for Wide-ResNet-50 v2, and 4096 for VGG-11 and VGG-19)
Each of such  sets of automatically computed  descriptors was combined with the clinical data and,  to avoid to  overwhelm the latter, the number of features in the former was reduced by a coarse  selection stage using the   univariate approach already described in section~\ref{subsec:FeatureSelectionClassifiers}.
Furthermore, we then applied the same  wrapper approach to investigate if the combination of automatic and clinical features had a degree of redundancy.
Straightforwardly, to avoid any bias all the operations described so far were performed respecting the training, validation and test split introduced before, and ensuring that the test was not used in any stage except for the final validation. 
 
Finally, the selected features were used to classify each patient in the two classes already mentioned, i.e. mild and severe, as reported in the last part of previous subsection, and using the same   learners already mentioned.

\subsection{End-to-end deep learning (DL) approach} \label{subsec:endDL}

The end-to-end DL approach was designed so that clinical information could influence the generation of useful features in image classification and vice-versa. Two different variants have been tested: in the first, CXR images were modified with the addition of an extra layer, in which pixels in fixed positions would code properly normalized clinical information, while the remaining pixels were filled with uniform white noise. The second variant consists in a multi-input network, which received separately CXR images and clinical information. Eventually, the latter variant proved to perform slightly better and it will be described in the following.

\begin{figure}[h]
\centering
\resizebox{0.99\columnwidth}{!}{
\includegraphics{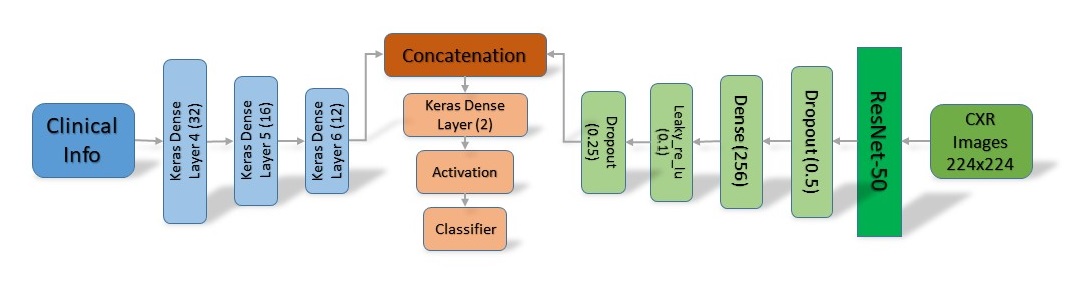}}
\caption{Workflow of the end-to-end deep learning approach.}
\label{fig:workflowdeeplearning}
\end{figure}

The architecture we adopted is composed by three main sections: one branch for each input accepts raw data and processes them to obtain a small number of relevant features, while a final common path concatenates the output features of the previous branches and uses them to provide the actual classification. A representation of the network can be found in \figurename~\ref{fig:workflowdeeplearning}.

Several different image classification networks have been tested (VGGs, ResNet, Inception and Xception variants) for the CXR branch. In general, smaller models appeared to perform better and the best results were observed with a ResNet50. The network has been adopted up to the last convolutional layer, while the final fully-connected layer and classification section have been removed. The number of generated output features has been reduced by a dropout layer with probability of 0.5, followed by a 256-neurons fully-connected layer, a leaky ReLU and a final dropout layer with probability 0.25.

All network tested were pre-trained on the ImageNet dataset, then on a publicly available repository of CXR images with the task of discriminating between healthy subjects and pneumonia patients \cite{mooney2020chest}; finally, the network was trained on the dataset presented here. 
We found that, with our architecture, data augmentation did not improve final classification performance and was, therefore, excluded from the processing pipeline. 
On the other hand, pre-training led to more consistent and slightly better results. 
As mentioned in section~\ref{subsec:imageStandardization}, the only image modification applied has been that of resizing all the input images to the size of 224 $\times$ 224, in order to preserve as much as possible, the features obtained with pre-training on the ImageNet dataset. 

The clinical information branch is a multi-layer perceptron (MLP): it is composed of a sequence of alternating fully-connected and non-linear layers; the adopted architecture consists of three fully-connected layers of decreasing size (32, 16 and 12 neurons), alternating with Rectified Linear Units (ReLUs). 

The common section of the network consists of a concatenation layer, which receives a total of 268 inputs (256 from the image branch and 12 from the clinical information branch) and feeds them to the actual classification section of the network (2-neurons fully-connected, softmax and classification layers).

In order to evaluate the impact of each data source, the model was trained as described above, as well as in two different versions modified to accept one data source only (i.e. changes consists in removal of one input branch and concatenation layer and with a change in the number of neurons in the final fully-connected layer). All versions underwent the same training procedure: a 20-epochs training phase with a SGD optimizer with a momentum set to 0.9. The weights used on the test set correspond with the iteration resulting in the lowest loss. The learning rate is fixed and equal to $10^{-4}$, while the batch size is set to 16. 


\begin{table}[th]
\centering
\caption{Best recognition performance attained by each of the learning methods when the experiments were executed according to the 10-fold cross-validation (20 repetitions). 
In the second column, ML and DL stands for Machine-Learning and Deep Learning, respectively. 
The last column reports the learners providing the results shown here. }
\resizebox{.99\textwidth}{!}{%
\begin{tabular}{@{}llcccl@{}}
\toprule
\textbf{Input data} & \textbf{Approach} & \textbf{Accuracy} & \textbf{Sensitivity} & \textbf{Specificity} & \textbf{Learner} \\ \midrule
\multirow{2}{*}{Clinical data}  & ML        & $.757 \pm .008$ & $.760 \pm .007 $ & $.754 \pm .011 $ & SVM \\
& DL & $.684\pm	.019$&$	.753\pm	.020$&	$.654\pm .012$ & MLP \\
\midrule
\multirow{3}{*}{CXR images}     & Handcrafted &  $.658 \pm .015$&$	.676\pm.016$ & $.638\pm.019$ &  LGR\\
& Hybrid & $.728\pm.038$&$	.769\pm	.072$&$	.680\pm	.076$  & VGG-11 + RF   \\
& End-to-end  & $.742\pm.010$&$	.748 \pm.019$&$	.738\pm	.013$   &    Resnet50 \\
\midrule
\multirow{3}{*}{\begin{minipage}{1in}Clinical data  and CXR images\end{minipage}} & Handcrafted & $.755\pm	.007$&$	.758\pm	.008$&$	.753\pm	.013$ &   SVM  \\
& Hybrid & $.769\pm	.054$&$	.788\pm	.064$&$	.747\pm	.059$    &  GoogleNet +  SVM   \\
& End-to-end  & $.748\pm	.008$&$	.745\pm	.017$&$.751\pm	.015$ &   Resnet50 + MLP  \\
\bottomrule
\end{tabular}}
\label{tab:ResultsBest10CV}
\end{table}

\begin{table}[th]
\centering
\caption{Best recognition performance attained by each of the learning methods when the experiments were executed according to the LOCO cross-validation. 
In the second column, ML and DL stands for Machine-Learning and Deep Learning, respectively. 
The last column reports the learners providing the results shown here. }
\resizebox{.99\textwidth}{!}{%
\begin{tabular}{@{}llcccl@{}}
\toprule
\textbf{Input data} & \textbf{Approach} & \textbf{Accuracy} & \textbf{Sensitivity} & \textbf{Specificity} & \textbf{Learner} \\ \midrule
\multirow{2}{*}{Clinical data}  & ML & $.734\pm	.044$&$	.699\pm	.158 $&$	.795\pm	.136$ & SVM \\
& DL & $.663\pm.016$&$	.709\pm	.032$&$	.644\pm	.018$ & MLP \\
\midrule
\multirow{3}{*}{CXR images}     & Handcrafted &  $.625\pm	.083$&$	.641\pm	.159$&$	.644\pm	.200$ & SVM\\
& Hybrid & $.693\pm.053$&$	.806\pm	.161	$&$.549\pm.213$  &   Vgg11 + SVM  \\
& End-to-end  & $.705\pm.010$&$	.720\pm	.011$&$	.696	\pm .015$  &  Resnet50   \\
\midrule
\multirow{3}{*}{\begin{minipage}{1in}Clinical data  and CXR images\end{minipage}} & Handcrafted & $.752\pm	.067$&$	.711\pm	.165$&$	.824\pm	.154$ &  LGR   \\
& Hybrid & $.743\pm	.061$&$	.769\pm	.189$&$	.685\pm	.155$    &  GoogleNet +  LGR   \\
& End-to-end  & $.709\pm.005$&$	.734\pm	.018$&$	.696	\pm.009$ &   Resnet50 + MLP  \\
\bottomrule
\end{tabular}}
\label{tab:ResultsBestCenter}
\end{table}

\section{Results}\label{sec:Result}

This section reports the results attained using the three approaches mentioned so far in staging the patients with COVID-19 in severe and mild classes.
The goal is to provide a   baseline  characterization of the performance achieved integrating together quantitative image data with clinical information by using state-of-the art approaches. 

\tablename~\ref{tab:ResultsBest10CV} and \ref{tab:ResultsBestCenter} present the best  recognition performance attained by each of  the learning methods when the experiments were executed according to the 10-fold and LOCO cross validation, respectively (see section~\ref{subsec:modelValidation} for further details).
In the former case, the results   are averaged over the 20 repetitions.
Furthermore, for the sake of readability  we omit  to report  the results achieved using the 3-fold cross validation since they are consistent with those performed in the 10-fold fashion.

The first two rows in both tables report the  performance in discriminating between patients with  mild and severe prognosis attained using  clinical data only.
In this respect, the row denoted by Machine Learning (ML)  shows the best performance achieved by the RFECV and by the learners described in the last part of section~\ref{subsec:handcrafted}, whereas the row denoted by Deep learning (DL) reports the performance returned by the multi-layer perceptron described in section~\ref{subsec:endDL}.
In the case of experiments performed in  10-fold cross validation (\tablename~\ref{tab:ResultsBest10CV}),  the best accuracy is up to 75.7\%, it is attained by an SVM retaining on average 11 clinical features, and the sensitivity and the specificity are almost balanced. 
This latter observation can be expected since the a-priori class distribution is not skewed. 
We also notice that the use of a deep network is sub-optimal in the classification task based on clinical information alone: this is likely due to the fact that, in contrast with the image case, pre-training of the network was impossible, due to the custom nature of input data. 
As a consequence, it is possible that the available number of samples was not sufficient to train the network to optimal performance.
The same observations hold also in the case of the experiments performed in a LOCO modality (\tablename~\ref{tab:ResultsBestCenter}), and  it is worth noticing the performance drops for both the ML and DL approaches. 
This can be due to the variation of data distribution among the centres, limiting the generalization capability of the learners.

In both \tablename~\ref{tab:ResultsBest10CV} and \ref{tab:ResultsBestCenter}, the   next two sections 
report the performance attained by the three methods described in section~\ref{sec:Methods} using only the CXR images and merging together the images with the clinical data, respectively.
With reference to the results reported in the section  ``CXR images'', they show that the use of   the images only  does not achieve the same performance obtained using the clinical data, whatever the method applied (\tablename~\ref{tab:ResultsBest10CV}).
Furthermore, the fact that the end-to-end DL has better results than the hybrid approach suggests that the fully connected portion of the CNN better exploits than a supervised classifier the information provided by the convolutional layers. 
In the case of the experiments performed in LOCO modality, there are still gaps with the results achieved using clinical data only, suggesting that all the learners suffer from the variability induced by the different centres.
Turning our attention to the results shown in section ``Clinical data and CXR images'', in the case of the experiments performed in 10 fold cross validation we notice that the integration between the two sources of information provides some benefits, permitting in some cases to improve the classification performance.  
Indeed, the hybrid approach achieves an accuracy up to  76.9\%, using the automatic features computed by the convolutional layers of the GoogleNet    and an SVM  classifier. 
The end-to-end DL approach slightly improves the performance with respect to the ones attained using only the images, suggesting that an approach fully based on DNN is not beneficial in this case, needing for further investigation.
In the case of the experiments run in LOCO mode we found that the integration of clinical data and CXR images is  beneficial 
as the largest accuracy is up to  75.2\%, with  improvements in terms of sensitivity and specificity. 

\section{Discussion} \label{sec:discussion}

This study originated during the first wave of  infection in Italy occurring in early spring, 2020, when thousands of people arrived every day in hospitals. 
Despite their apparently similar conditions, 
some lived the infection as a seasonal flu while others rapidly deteriorated, making intensive care necessary. 
This situation is common worldwide and, to fight the pandemic, in the last months the whole scientific community   has carried out  relevant research efforts in different fields of knowledge.

Artificial intelligence is one of the scientific disciplines that has been attracting more attention, offering the possibility to process and extract knowledge and insights from the massive amount of data generated during the pandemic, and it  has mostly impacted prediction, diagnosis and treatment.
Within this context, large efforts have been directed towards the analysis of radiological images and, according to the analysis presented by Greenspan et.al. \cite{bib:greenspan2020position},  detection of COVID-19 pneumonia in both  CT and CXR \cite{bib:zhang2020clinically,bib:minaee2020deep} is the field where large research has been directed to. 
Recently, there has been  growing interest in the development of  AI models to predict the severity of the COVID-19 infections    because of the pressure on the  hospitals, where even during the second pandemic wave we have assisted to    an increasing demand for beds in both ordinary wards and    intensive care units.  
 The few papers  available in this field use CT images, but  several guidelines and statement do not encourage the use of CT over CXR \cite{rubin2020others} and for several practice reasons  CXR imaging is used due to the difficulty of moving bedridden patients,    the lack of CT machine slots,  the risk of cross-infection, etc..

To deal with this issue, here we have presented a multicentre retrospective study where we collected the CXR examination and clinical data of 820 patients from 6 Italian hospitals. 
Furthermore, we have  investigated different AI approaches to provide  to researchers and practitioners a baseline performance reference to foster further studies.

\begin{figure}[h]
\centering
\resizebox{0.99\textwidth}{!}{
\includegraphics{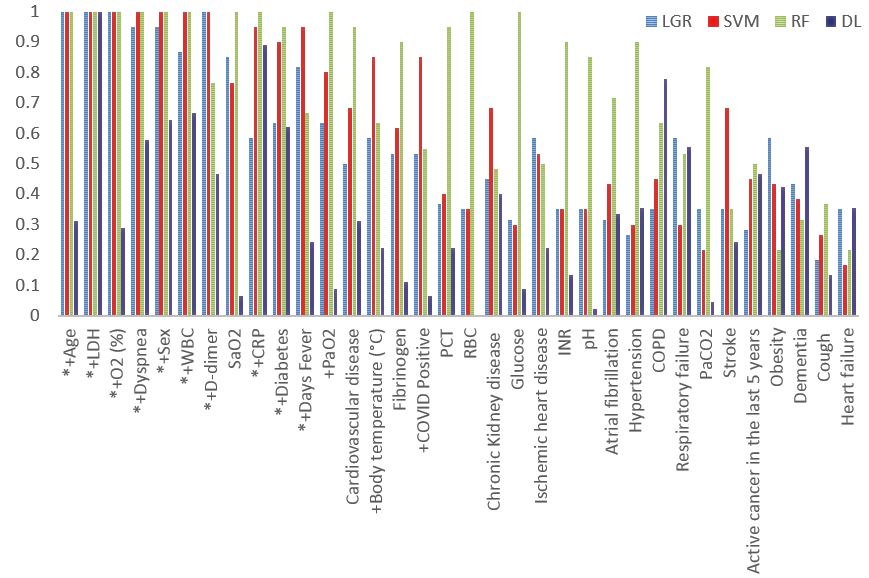}}
\caption{Clinical feature importance represented by the rate each descriptor was selected by the RFECV wrapper during both the 10-fold and LOCO cross validation experiments using the three classifiers (LGR, SVM and RF series). The DL series represents feature importance estimated as the maximum absolute value of weights in the first layer of the perceptron of the DL network, after averaging over folds and repetitions and rescaling in the [0,1] interval. Moreover, the ``*'' or a ``+''    reported before each  feature name means that  it is included in the feature set used to get the best handcrafted results reported in the first section of  \tablename~\ref{tab:ResultsBest10CV} and \ref{tab:ResultsBestCenter}, respectively.}
\label{fig:percentageChosenClinical10fold}
\end{figure}

With reference to the   population characteristics, we found interesting reports on the  age and gender distribution.
Women were both less and older, suggesting that they become less ill and suffer from more serious conditions at an older age than men; also the women mortality was lower, as 72\% were male confirming the male mortality reported in China (73\%) by Chen et.al.  \cite{bib:chen2020clinical}.
The male-related susceptibility and the higher male-mortality rate was also reported by Borges et.al. \cite{bib:borges2020novel}, who  analyses the data of 59,254 patients from 11 different countries.
The second main finding was that 87\% of patients had at least one comorbidity (\figurename~\ref{fig:diseasedistribution}), suggesting that, in most cases, the  conditions leading to hospitalisation occur in  patients with coexisting disorders. 
The most common disease (in 45 \% of cases) was hypertension, confirming the results reported by Yang et.al. \cite{bib:yang2020prevalence}, who  meta-analysed the data of 1,576 infected patients from seven studies and reported an hypertension prevalence of 21\%.

Let us now turn the attention to the results attained by the AI approaches that process  the clinical data only. 
Using  a normalized unitary scale,~\figurename~\ref{fig:percentageChosenClinical10fold} shows the rate each clinical descriptor was included in the selected feature subset by the RFECV wrapper, distinguishing also per classifier used. The figure shows the cumulative results observed running  both the  10-fold and LOCO cross validation experiments. 
We opted for this cumulative representation since the trend is very similar in both the experiments. 
Furthermore, the readers can find in the figure also the set of biomarkers providing the best performance shown in the first section of \tablename~\ref{tab:ResultsBest10CV} and \ref{tab:ResultsBestCenter}, which are denoted by reporting before an ``*'' or a ``+'' for 10-fold and LOCO cross validation experiments, respectively.
Interestingly, \figurename~\ref{fig:percentageChosenClinical10fold} shows that  age, LDH and $O_2$, were chosen in every fold for all the  classifiers. 
If we  used only such three descriptors,   the average classification accuracy attained by  learners  in 10-fold and LOCO cross validation is  equal to $0.74 \pm 0.05 $ and to $0.70 \pm 0.10 $, respectively. 
Moreover,  sex, dyspnoea and WBC were always selected by the wrapper with the SVM and RF,  whereas the D-dimer was always selected by the logistic regressor and by SVM.
Oppositely, heart failure and cough were scarcely selected. 
Notably, some features such as LDH, D-dimer and SaO$_2$ were selected very frequently despite a high fraction of data was obtained by imputation (see \tablename~\ref{tab:databaseinformation}). 
We deem that is mostly  related to the strong differences in the distributions of these features between the two classes.   


~\figurename~\ref{fig:percentageChosenClinical10fold} also shows in dark blue the feature relevance estimated by the deep learning approach (DL series).
In this case the feature relevance was estimated as the maximum across neurons of the absolute value of the weights in the perceptron first layer. Results have been averaged over cross-validation folds and repetitions and rescaled to the [0,1] interval in order to match the other three series. Comparing the results with those obtained with the RFECV wrapper, it is clear that the only feature with the maximum relevance for all approaches is LDH, while sex, dyspnoea, WBC and  CRP present a score higher than 0.5 in all series. 
The impact of the other clinical attributes appears to vary significantly depending on the adopted approach.
For example, a high value of D-dimer and WBC have shown to be an important risk factor for negative outcome \cite{bib:zhang2020d, bib:petrilli2020factors, bib:henry2020hematologic}. Furthermore, D-dimer, WBC and other clinical features like dyspnoea and LDH are indicators of pulmonary compromise, infection, tissue damage \cite{bib:li2020elevated} and a pro-thrombotic state \cite{bib:naymagon2020admission} respectively. Finally, from our first statistical analysis (section \ref{sec:StatisticalAnalysis}) of the dataset and from the result in shown in ~\figurename~\ref{fig:percentageChosenClinical10fold}   the patient gender showed to have an in important role for classifying the patient severity.
The reasons behind this difference appear to be related to the stark difference in immune system responses, with females causing stronger immune responses to pathogens. This difference can be a major contributing factor to viral load, disease severity, and mortality. Furthermore, differences in sex hormone environments could also be a determinant of viral infections as oestrogen has immune-stimulating effects while testosterone has immune-suppressive effects \cite{bib:pradhan2020sex}.

With reference to the results attained by the  handcrafted approach, we found that the best results in terms of accuracy are statistically lower than those attained by the clinical descriptors ($p<$ 0.001 and $p<$ 0.05 for 10-fold and LOCO cross validation, respectively). 

The approach that computes handcrafted features from the images also unfavourably compares with those using CNNs.
Indeed, comparing with the hybrid and the end-to-end DL approaches we found that the performances are statistically different in both the 10-fold and LOCO cross validation tests, as we always got $p <$ 0.05. 
No statistically different performances were found, instead, between the end-to-end and hybrid approaches.

\begin{figure}[H]
  \centering
  \subfloat[Handcrafted approach]
  {\includegraphics[width=0.99\textwidth]{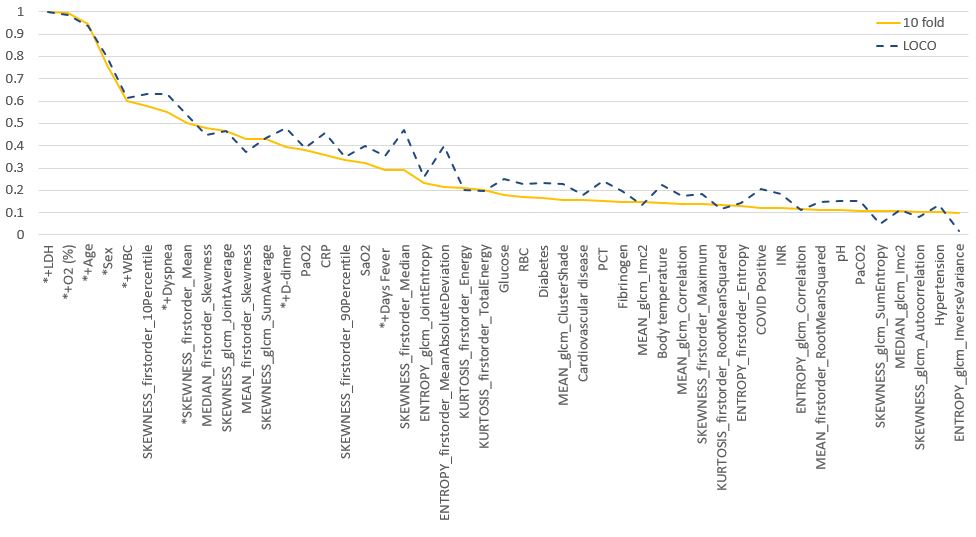}
  \label{fig:percentageChosenRadAndClinical}}
  \vspace{0.5em}
  \subfloat[Hybrid approach]
  {\includegraphics[width=0.9\textwidth]{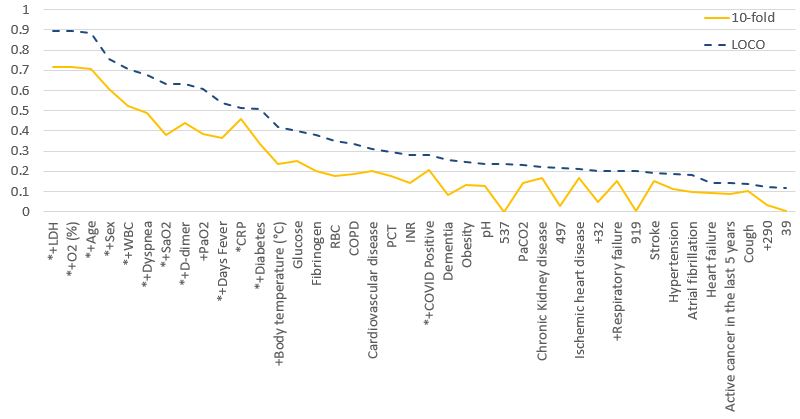}
  \label{fig:percentageChosenDeepAndClinical_googlenet}}
  \caption{Importance of clinical and handcrafted (panel A) or automatically learnt  features (panel B)   measured as  the rate each descriptor was selected by the RFECV wrapper during the  10-fold and LOCO cross-validation experiments considering all the the  three classifiers employed. The y axis scale is normalized to one.
  Moreover, we add a ``*'' or a ``+''     before each  feature name if  it is included in the feature set used to get the best handcrafted or hybrid results reported in the last section of  \tablename~\ref{tab:ResultsBest10CV} and \ref{tab:ResultsBestCenter}, respectively. }
\end{figure}

Furthermore, \figurename~\ref{fig:percentageChosenRadAndClinical} shows the feature importance of the 40 most selected handcrafted descriptors by the RFECV wrapper during the experiments in 10-fold and LOCO cross-validation. 
The feature relevance is computed as the number of times a feature is included in the selected subset during all the experiments performed using all the learners and, for the sake of clarity, all the values are normalized in [0,1].
The plot shows that the top-five descriptors most frequently detected as discriminative are clinical measures, followed by several texture measures almost equally distributed between the first- and second-order measures.  
For the sake of completeness, in this figure on the x-axis we add  a ``*'' or a ``+''    before each  feature name when it is included in the feature set used to get the best  results by combining handcrafted measures from CXR images and clinical data,  reported in the last section of  \tablename~\ref{tab:ResultsBest10CV} and \ref{tab:ResultsBestCenter}, respectively. 

\begin{figure}[H]
  \centering
  \subfloat[Handcrafted approach]
  {\includegraphics[width=0.7\textwidth]{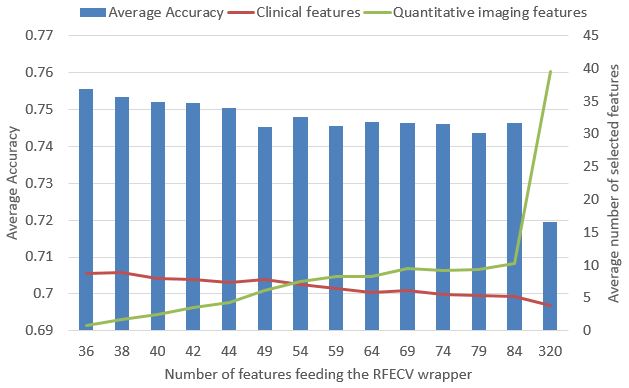}
  \label{fig:accuarcyClinicalRadiomiFeatures}}
  \vspace{0.5em}
  \subfloat[Hybrid approach]
{\includegraphics[width=0.7\textwidth]{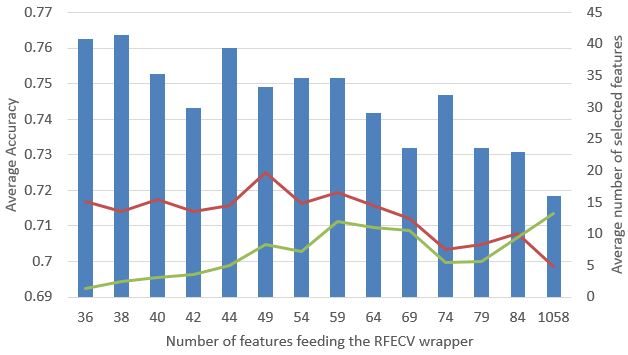}
  \label{fig:accuarcyClinicalDeepFeatures}}
  \caption{Variation of the average classification accuracy (blue bars) with the number of   features feeding the RFECV wrapper. 
The red and green curves show the number of clinical and texture  features selected  by  the RFECV wrapper, respectively. 
The experiments plotted here refer to the best results shown in \tablename~\ref{tab:ResultsBest10CV} integrating clinical and imaging features for the handcrafted (panel A) and hybrid approach (panel B).}

\end{figure}

We now  analyse how the performance of the handcrafted approach vary with the number of features selected by the coarse step, which fed the fine selection based on the RFECV method, as described in section \ref{subsec:handcrafted}.  
To this end,   \figurename~\ref{fig:accuarcyClinicalRadiomiFeatures}  reports on  the x-axis  the number of features in input to the RFECV, which ranges from 36 (i.e. 34 clinical plus 2 texture measures)  up to 84 (i.e. 34 clinical plus 50 texture measures), plus the last value where the RFECV received all the clinical and all the image features\footnote{The experiments plotted in \figurename~\ref{fig:accuarcyClinicalRadiomiFeatures} refer to the best results shown in \tablename~\ref{tab:ResultsBest10CV} integrating clinical and imaging features by the handcrafted approach.}.
The bars show the average classification accuracy (y-axis, left side), while the  curves in red and green show the average number of clinical and handcrafted texture features selected by the RFECV, respectively (y-axis, right side).
As already noticed in \tablename~\ref{tab:ResultsBest10CV}, the use of texture measures does not improve the performance attained using the clinical descriptors; this is also confirmed by 
observing that, as the number of input features increases,  the wrapper tends to select more imaging biomarkers than clinical ones, dropping the performance. 
This may remark the importance of using both  clinical and imaging biomarkers since they may provide complementary information: while the former, and especially comorbidities, refers to the functional reserve of the patient, the latter   may quantify the actual impact on the lungs. 
Indeed, fit patients with severe infection and damage are as likely as unfit-patients with less severe infections to have a poor prognosis. 
Although not reported, similar considerations can be derived   in the case of LOCO cross-validation where we   noticed that the best performance are attained by an almost balanced number of  clinical and imaging features. 

\begin{figure}
        \begin{subfigure}[b]{0.25\textwidth}
                \centering
                \includegraphics[width=.8\linewidth]{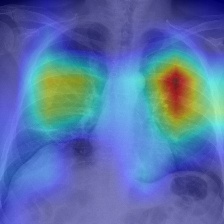}
                \caption{Mild class, all neurons}
                \label{fig:MildClassALLNeurons}
        \end{subfigure}%
        \begin{subfigure}[b]{0.25\textwidth}
                \centering
                \includegraphics[width=.8\linewidth]{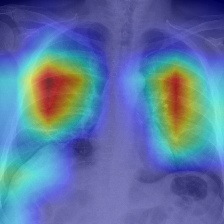}
                \caption{Mild class, 40 most selected neurons}
                \label{fig:MildClass40Neurons}
        \end{subfigure}%
        \begin{subfigure}[b]{0.25\textwidth}
                \centering
                \includegraphics[width=.8\linewidth]{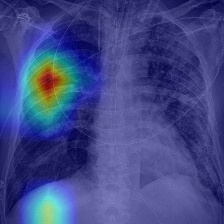}
                \caption{Severe class, all neurons}
                \label{fig:SevereClassALLNeurons}
        \end{subfigure}%
        \begin{subfigure}[b]{0.25\textwidth}
                \centering
                \includegraphics[width=.8\linewidth]{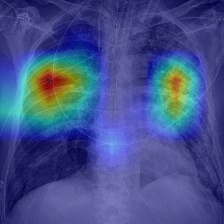}
                \caption{Severe class, 40 most selected neurons}
                \label{fig:SevereClass40Neurons}
        \end{subfigure}
        \caption{Two examples of the activation maps provided by the Grad-CAM approach, using all the neurons in the dense layer of the CNN dense layer or all the 40  neurons selected by the RFECV wrapper. }
        \label{fig:GradCamHybrid}
\end{figure}

With reference to the results attained by the hybrid approach on the CXR images only, we found that the best results are statistically lower than those attained by the clinical descriptors for 10-fold cross validation ($p < $ 0.001) but no differences were found with LOCO cross validation ($p =  0.24$). Among the three learners used with the hybrid approach, the best results with 10-fold cross validation are obtained with RF ($p < $ 0.001, Kruskal-Wallis and Dunn's test) while no differences were found with LOCO validation. Furthermore, comparing with a full DL approach, the hybrid provide lower performance ($p < $ 0.05 and $p = $ 0.24 for 10-fold and LOCO cross validation, respectively), suggesting that a fully connected layer better exploits the automatic features computed by the convolutional layers of the CNNs. 
As in \figurename~\ref{fig:percentageChosenRadAndClinical}, we show in \figurename~\ref{fig:percentageChosenDeepAndClinical_googlenet}  the feature importance of the  40 most selected descriptors by the RFECV wrapper during the experiments in 10-fold and LOCO cross-validation using the GoogleNet. 
The plot shows that the features most frequently detected as discriminative are clinical measures with some neurons of the dense layer  that, although few in number, permits to improve the classification accuracy.  
To deepen the results,  \figurename~\ref{fig:GradCamHybrid} shows how much the selected neurons contribute to the network   predictions. 
To this goal, we first depict the regions of input that are important for outputs provided by the CNN (panels a and c in the figure) by applying the  Gradient-weighted Class Activation Mapping (Grad-CAM) approach~\cite{GradCAM}.
In a nutshell, Grad-CAM uses the class-specific gradient information flowing into the final convolutional layer   to produce a coarse localization map of the relevant regions in the image.
Next, we ran the same algorithm using only the 40 neurons in the dense layer that were mostly selected by the RFECV wrapper (panels b and d in the same figure).
The  visually inspection of the figure shows that the regions activated by the 40 neurons cover most of the areas activated by the whole dense layer, confirming that the wrapper correctly identifies the neurons carrying most of the information. 
Finally, as in~\figurename~\ref{fig:accuarcyClinicalRadiomiFeatures}, \figurename~\ref{fig:accuarcyClinicalDeepFeatures} shows that    using  all the features automatically learnt does not help the learner improving the accuracy, whilst a limited and small number of descriptors is beneficial.

The end-to-end deep learning approach was built with the intuition that, through joined training of clinical information and images, it would be possible to generate better features for classification than by using either source alone. 
This idea was at least partly vindicated, as the classification results for the fully-DL approach proved higher for the combined approach than from either single source in the 10-fold cross-validation scenario ($p =  0.02$) and, arguably, for the LOCO case, as well ($p =  0.06$).

Furthermore, as already mentioned, classification accuracy from images alone is better than other methods, confirming the well-established finding that CNNs are powerful approaches for image classification. 
Oppositely, a neural network-based  approach suffered particularly in achieving good performance with clinical information as inputs. 
The most likely cause for this under-performance is the fact that the clinical information structure is not standard and, therefore, it was impossible to adopt already tested network models and, more importantly, to pre-train the network on other datasets. 
It is likely that further fine-tuning of the design and training procedure of the custom multi-layer perceptron adopted for clinical-info classification could further improve results both with this specific input source, as well as for the combined model. 
A similar result is expected with an increase in size of the available dataset, as this section of the network did not undergo any pre-training, as mentioned above.

This study has also some limitations. 
First, patient enrolment was not globally randomized but instead conducted to populate the two classes with a roughly homogeneous number of cases. This implies that training data do not reflect the true a-priori probabilities of the target classes. 
On the other hand, sampling within the mild and severe classes is unbiased because patients were randomly enrolled. 

Although this may bias the estimate of classification accuracy, 
there exist methods for adjusting the outputs of a trained classifier with respect to different  prior probabilities without having to retrain the model, even when these probabilities are not known in advance \cite{latinne2001adjusting}.

A further limitation but, from another point of view a key feature, of the study is the lack of full standardization of images and clinical data in the dataset.
The dataset was built during spring 2020 when Italy was under lockdown and Italian hospitals and doctors were under pressure due to the huge amount of patients requiring hospitalization. 
Under these circumstances, full standardization of clinical data collection and images acquisition could not be achieved, and we decided to collect CXR images gathered under any conditions and all the clinical data most commonly acquired at the time of patients hospitalization. 
This led to a dataset that reflects these circumstances with many missing values among clinical data and images acquired with unstandardized clinical protocol (i.e. patient position and breath holding) and various devices.
Although on the one side this may   represent a limitation, on the other side it may be  an advantage because this dataset could challenge the AI community on real data collected under critical circumstances.
Another limitation may be the ever-changing landscape of the pandemic. 
Compared to the first wave, in many countries, and especially in Europe, the second wave has been characterised by younger patients with early symptoms in the emergency department.
This may suggest to   periodically re-train the learners to follows the disease evolution, or to investigate the use methods able to cope with concept drifts~\cite{bib:ConceptDrift}.  

\subsection{Take-home messages and future works}

In this preliminary analysis the use of image-derived data provide reduced  predictive performance improvement with respect to the use of clinical data alone.
The analysis of clinical data, instead, showed that a number of measures have robust predictive potential. 
Clinical data such as Age, LDH, O$_2$, Dyspena, Sex, WBC, D-dimer, SaO$_2$   are consistently selected across the different validation condition and classifiers tested in this work, representing a set of biomarkers that can have impact in the clinical practice helping physicians and care-managers planning the bed allocations.

The poor standardization of images in the dataset could be a possible cause of the results attained here, as it has led to a classification problem hard  to be  addressed by the  tested approaches.
Indeed, beyond the variability introduced by non-standardized acquisition conditions such as patient positions and imaging device,  the number of various medical devices, metal objects and other artefacts (e.g. pacemakers, catheters, prosthesis, etc.) that can be observed within the field of view are additional sources of difficulty for the learners. 
To further  explore the dataset, this suggests using methods that can manage such variability, for instance by  disregarding those images not meeting some quality criterion that can be learnt in parallel with the classification task.
 With reference to the approaches investigated here, deepening how data augmentation impact  network training, performing ablation studies on the hybrid approach  as well as on  network sizes for the end-to-end DL procedure are future directions of investigation. 
Furthermore,  to improve the quality of DNNs we deem that joint learning could be another direction of investigation, enabling the possibility  to extract correlated information across clinical and imaging data to the used to enforce  the network weights to be shared across these networks. 

In conclusion, the dataset presented here is unique, offering a large number of CXR for prognostic purposes, placing side by side with similar efforts that use CT images \cite{bib:chassagnon2020ai}. 
While this repository 
lets the machine learning community to challenge their methods with poorly standardized data,   the efforts to collect  a large repository  cannot be afforded by such  community, asking for   the collaboration of researchers from different backgrounds, clinicians, and institutes.
Furthermore, the quantitative results reported offer a preliminary evaluation of the prognostic performance attainable using AI approaches spanning from the use of handcrafted image descriptors to a fully automatic approach based on DNNs.
The use of AI in this domain can open the chance to develop fast and low-cost clinical protocols.

\section*{Data availability}
The dataset generated and analysed in this study is publicly available to members of the   scientific community upon request at    {\em aiforcovid.radiomica.it} 
Beyond that, we encourage  other hospitals and clinical centres to join the network to share their data; in this case, contacts for data sharing are also available on the website.  
As mentioned in section~\ref{sec:materials}, the dataset contains the CXR images, the clinical data listed in \tablename~\ref{tab:databaseinformation}, the labels, the blind association between each image and the acquisition centre, and the acquisition information.
 The manual segmentation masks mentioned in section~\ref{subsec:handcrafted} are not publicly available at the time this manuscript is submitted, and they will be added later on.

 \section*{Acknowledgments}
The authors wish to thank Amazon Web Services (AWS) and the AWS Diagnostic Development Initiative for the support in putting in place the data management infrastructure.

\bibliographystyle{unsrt}  
\bibliography{arxis}  

\begin{thebibliography}{10}

\bibitem{bib:schiaffino2020diagnostic}
Simone Schiaffino, Stefania Tritella, Andrea Cozzi, Serena Carriero, Lorenzo
  Blandi, Laurenzia Ferraris, and Francesco Sardanelli.
\newblock Diagnostic performance of {Chest X-ray} for {COVID-19} pneumonia
  during the {SARS-CoV-2} pandemic in {Lombardy}, {Italy}.
\newblock {\em Journal of thoracic imaging}, 35(4):W105--W106, 2020.

\bibitem{bib:ai2020correlation}
Tao Ai, Zhenlu Yang, Hongyan Hou, Chenao Zhan, Chong Chen, Wenzhi Lv, Qian Tao,
  Ziyong Sun, and Liming Xia.
\newblock Correlation of {Chest}{CT} and {RT-PCR} testing in coronavirus
  disease 2019 ({COVID-19}) in china: a report of 1014 cases.
\newblock {\em Radiology}, page 200642, 2020.

\bibitem{bib:lu2020outbreak}
Hongzhou Lu, Charles~W Stratton, and Yi-Wei Tang.
\newblock Outbreak of pneumonia of unknown etiology in {Wuhan}, {China}: The
  mystery and the miracle.
\newblock {\em Journal of medical virology}, 92(4):401--402, 2020.

\bibitem{bib:ACR2020}
{American College of Radiology}.
\newblock {ACR Recommendations for the use of Chest Radiography and Computed
  Tomography (CT) for Suspected COVID-19 Infection}.
\newblock
  \url{https://www.acr.org/Advocacy-and-Economics/ACR-Position-Statements/Recommendations-for-Chest-Radiography-and-CT-for-Suspected-COVID19-Infection}.
\newblock Online; accessed November, 30 2020.

\bibitem{vancheri2020radiographic}
Sergio~Giuseppe Vancheri, Giovanni Savietto, Francesco Ballati, Alessia Maggi,
  Costanza Canino, Chandra Bortolotto, Adele Valentini, Roberto Dore,
  Giulia~Maria Stella, Angelo~Guido Corsico, et~al.
\newblock Radiographic findings in 240 patients with {COVID-19} pneumonia:
  time-dependence after the onset of symptoms.
\newblock {\em European Radiology}, page~1, 2020.

\bibitem{bib:zanardo2020bringing}
Moreno Zanardo, Simone Schiaffino, and Francesco Sardanelli.
\newblock Bringing radiology to patient's home using mobile equipment: A weapon
  to fight {COVID-19} pandemic.
\newblock {\em Clinical Imaging}, 68:99--101, 2020.

\bibitem{bib:liu2017tx}
Chang Liu, Yu~Cao, Marlon Alcantara, Benyuan Liu, Maria Brunette, Jesus
  Peinado, and Walter Curioso.
\newblock Tx-cnn: Detecting tuberculosis in {Chest X-ray} images using
  convolutional neural network.
\newblock In {\em 2017 IEEE International Conference on Image Processing
  (ICIP)}, pages 2314--2318. IEEE, 2017.

\bibitem{bib:yan2019combining}
Fengqi Yan, Xin Huang, Yao Yao, Mingming Lu, and Maozhen Li.
\newblock Combining lstm and densenet for automatic annotation and
  classification of chest x-ray images.
\newblock {\em {IEEE Access}}, 7:74181--74189, 2019.

\bibitem{bib:challenge2018radiological}
{Radiological Society of North America}.
\newblock {RSNA Pneumonia Detection Challenge (2018)}.
\newblock
  \url{https://www.rsna.org/en/education/ai-resources-and-training/ai-image-challenge/RSNA-Pneumonia-Detection-Challenge-2018}.
\newblock Online; accessed 15 November 2020.

\bibitem{bib:xu2018deepcxray}
Xiuyuan Xu, Quan Guo, Jixiang Guo, and Zhang Yi.
\newblock Deepcxray: Automatically diagnosing diseases on {Chest X-rays} using
  deep neural networks.
\newblock {\em {IEEE Access}}, 6:66972--66983, 2018.

\bibitem{bib:greenspan2020position}
Hayit Greenspan, Ra{\'u}l San~Jos{\'e} Est{\'e}par, Wiro~J Niessen, Eliot
  Siegel, and Mads Nielsen.
\newblock Position paper on {COVID-19} imaging and {AI}: From the clinical
  needs and technological challenges to initial {AI} solutions at the lab and
  national level towards a new era for {AI} in healthcare.
\newblock {\em Medical Image Analysis}, 66:101800, 2020.

\bibitem{bib:zhang2020clinically}
Kang Zhang, Xiaohong Liu, Jun Shen, Zhihuan Li, Ye~Sang, Xingwang Wu, Yunfei
  Zha, Wenhua Liang, Chengdi Wang, Ke~Wang, et~al.
\newblock Clinically applicable {AI} system for accurate diagnosis,
  quantitative measurements, and prognosis of {COVID-19} pneumonia using
  computed tomography.
\newblock {\em Cell}, 2020.

\bibitem{bib:minaee2020deep}
Shervin Minaee, Rahele Kafieh, Milan Sonka, Shakib Yazdani, and
  Ghazaleh~Jamalipour Soufi.
\newblock {Deep-COVID: Predicting COVID-19 from Chest X-Ray Images Using Deep
  Transfer Learning}.
\newblock {\em Medical Image Analysis}, 2020.

\bibitem{bib:gozes2020rapid}
Ophir Gozes, Maayan Frid-Adar, Hayit Greenspan, Patrick~D Browning, Huangqi
  Zhang, Wenbin Ji, Adam Bernheim, and Eliot Siegel.
\newblock Rapid {AI} development cycle for the coronavirus ({COVID-19})
  pandemic: Initial results for automated detection \& patient monitoring using
  deep learning {CT} image analysis.
\newblock {\em arXiv preprint arXiv:2003.05037}, 2020.

\bibitem{bib:wynants2020prediction}
Laure Wynants, Ben Van~Calster, Marc~MJ Bonten, Gary~S Collins, Thomas~PA
  Debray, Maarten De~Vos, Maria~C Haller, Georg Heinze, Karel~GM Moons,
  Richard~D Riley, et~al.
\newblock Prediction models for diagnosis and prognosis of {COVID-19}
  infection: systematic review and critical appraisal.
\newblock {\em British Medical Journal}, 369, 2020.

\bibitem{bib:moons2019probast}
Karel~GM Moons, Robert~F Wolff, Richard~D Riley, Penny~F Whiting, Marie
  Westwood, Gary~S Collins, Johannes~B Reitsma, Jos Kleijnen, and Sue Mallett.
\newblock Probast: a tool to assess risk of bias and applicability of
  prediction model studies: explanation and elaboration.
\newblock {\em Annals of internal medicine}, 170(1):W1--W33, 2019.

\bibitem{bib:yue2020machine}
Hongmei Yue, Qian Yu, Chuan Liu, Yifei Huang, Zicheng Jiang, Chuxiao Shao,
  Hongguang Zhang, Baoyi Ma, Yuancheng Wang, Guanghang Xie, et~al.
\newblock Machine learning-based {CT} radiomics method for predicting hospital
  stay in patients with pneumonia associated with {SARS-CoV-2} infection: a
  multicenter study.
\newblock {\em Annals of translational medicine}, 8(14), 2020.

\bibitem{bib:chassagnon2020ai}
Guillaume Chassagnon, Maria Vakalopoulou, Enzo Battistella, Stergios
  Christodoulidis, Trieu-Nghi Hoang-Thi, Severine Dangeard, Eric Deutsch,
  Fabrice Andre, Enora Guillo, Nara Halm, et~al.
\newblock {AI-Driven quantification, staging and outcome prediction of COVID-19
  pneumonia}.
\newblock {\em Medical Image Analysis}, page 101860, 2020.

\bibitem{bib:leeuwenberg2020prediction}
Artuur~M Leeuwenberg and Ewoud Schuit.
\newblock Prediction models for {COVID-19} clinical decision making.
\newblock {\em The Lancet Digital Health}, 2(10):e496--e497, 2020.

\bibitem{bib:yang2020clinical}
Xiaobo Yang, Yuan Yu, Jiqian Xu, Huaqing Shu, Hong Liu, Yongran Wu, Lu~Zhang,
  Zhui Yu, Minghao Fang, Ting Yu, et~al.
\newblock {Clinical course and outcomes of critically ill patients with
  SARS-CoV-2 pneumonia in Wuhan, China: a single-centered, retrospective,
  observational study}.
\newblock {\em The Lancet Respiratory Medicine}, 2020.

\bibitem{bib:lungseg}
{Imlab-UIIP}.
\newblock {Lung Segmentation (2D)}.
\newblock \url{https://github.com/imlab-uiip/lung-segmentation-2dR}.
\newblock Online; accessed 19 October 2020.

\bibitem{bib:jaeger2014two}
Stefan Jaeger, Sema Candemir, Sameer Antani, Y{\`\i}-Xi{\'a}ng~J W{\'a}ng,
  Pu-Xuan Lu, and George Thoma.
\newblock Two public {Chest X-ray} datasets for computer-aided screening of
  pulmonary diseases.
\newblock {\em Quantitative imaging in medicine and surgery}, 4(6):475, 2014.

\bibitem{bib:shiraishi2000development}
Junji Shiraishi, Shigehiko Katsuragawa, Junpei Ikezoe, Tsuneo Matsumoto,
  Takeshi Kobayashi, Ken-ichi Komatsu, Mitate Matsui, Hiroshi Fujita, Yoshie
  Kodera, and Kunio Doi.
\newblock Development of a digital image database for {Chest}radiographs with
  and without a lung nodule: receiver operating characteristic analysis of
  radiologists' detection of pulmonary nodules.
\newblock {\em American Journal of Roentgenology}, 174(1):71--74, 2000.

\bibitem{bib:mutualinfocontinuousvar}
Brian Ross.
\newblock Mutual information between discrete and continuous data sets.
\newblock {\em PloS one}, 9:e87357, 02 2014.

\bibitem{bib:guyon2002gene}
Isabelle Guyon, Jason Weston, Stephen Barnhill, and Vladimir Vapnik.
\newblock Gene selection for cancer classification using support vector
  machines.
\newblock {\em Machine learning}, 46(1-3):389--422, 2002.

\bibitem{arcuri2013parameter}
Andrea Arcuri and Gordon Fraser.
\newblock Parameter tuning or default values? an empirical investigation in
  search-based software engineering.
\newblock {\em Empirical Software Engineering}, 18(3):594--623, 2013.

\bibitem{bib:penny2011statistical}
William~D Penny, Karl~J Friston, John~T Ashburner, Stefan~J Kiebel, and
  Thomas~E Nichols.
\newblock {\em Statistical parametric mapping: the analysis of functional brain
  images}.
\newblock Elsevier, 2011.

\bibitem{bib:haralick1973textural}
Robert~M Haralick, Karthikeyan Shanmugam, and Its'~Hak Dinstein.
\newblock Textural features for image classification.
\newblock {\em {IEEE Transactions on systems, man, and cybernetics}},
  (6):610--621, 1973.

\bibitem{bib:krizhevsky2014one}
Alex Krizhevsky.
\newblock One weird trick for parallelizing convolutional neural networks.
\newblock {\em arXiv preprint arXiv:1404.5997}, 2014.

\bibitem{bib:simonyan2014very}
Karen Simonyan and Andrew Zisserman.
\newblock Very deep convolutional networks for large-scale image recognition.
\newblock {\em arXiv preprint arXiv:1409.1556}, 2014.

\bibitem{bib:he2016deep}
Kaiming He, Xiangyu Zhang, Shaoqing Ren, and Jian Sun.
\newblock Deep residual learning for image recognition.
\newblock In {\em Proceedings of the IEEE conference on computer vision and
  pattern recognition}, pages 770--778, 2016.

\bibitem{bib:xie2017aggregated}
Saining Xie, Ross Girshick, Piotr Doll{\'a}r, Zhuowen Tu, and Kaiming He.
\newblock Aggregated residual transformations for deep neural networks.
\newblock In {\em Proceedings of the IEEE conference on computer vision and
  pattern recognition}, pages 1492--1500, 2017.

\bibitem{bib:zagoruyko2016wide}
Sergey Zagoruyko and Nikos Komodakis.
\newblock Wide residual networks.
\newblock {\em arXiv preprint arXiv:1605.07146}, 2016.

\bibitem{bib:iandola2016squeezenet}
Forrest~N Iandola, Song Han, Matthew~W Moskewicz, Khalid Ashraf, William~J
  Dally, and Kurt Keutzer.
\newblock Squeezenet: Alexnet-level accuracy with 50x fewer parameters and
  \textless{} 0.5 mb model size.
\newblock {\em arXiv preprint arXiv:1602.07360}, 2016.

\bibitem{bib:huang2017densely}
Gao Huang, Zhuang Liu, Laurens Van Der~Maaten, and Kilian~Q Weinberger.
\newblock Densely connected convolutional networks.
\newblock In {\em Proceedings of the IEEE conference on computer vision and
  pattern recognition}, pages 4700--4708, 2017.

\bibitem{bib:szegedy2015going}
Christian Szegedy, Wei Liu, Yangqing Jia, Pierre Sermanet, Scott Reed, Dragomir
  Anguelov, Dumitru Erhan, Vincent Vanhoucke, and Andrew Rabinovich.
\newblock Going deeper with convolutions.
\newblock In {\em Proceedings of the IEEE conference on computer vision and
  pattern recognition}, pages 1--9, 2015.

\bibitem{bib:ma2018shufflenet}
Ningning Ma, Xiangyu Zhang, Hai-Tao Zheng, and Jian Sun.
\newblock Shufflenet v2: Practical guidelines for efficient {CNN} architecture
  design.
\newblock In {\em Proceedings of the European conference on computer vision
  (ECCV)}, pages 116--131, 2018.

\bibitem{bib:sandler2018mobilenetv2}
Mark Sandler, Andrew Howard, Menglong Zhu, Andrey Zhmoginov, and Liang-Chieh
  Chen.
\newblock Mobilenetv2: Inverted residuals and linear bottlenecks.
\newblock In {\em Proceedings of the IEEE conference on computer vision and
  pattern recognition}, pages 4510--4520, 2018.

\bibitem{mooney2020chest}
{Mooney, Paul}.
\newblock {Chest X-ray images (Pneumonia). 2017}.
\newblock \url{https://www.kaggle.com/paultimothymooney/chest-xray-pneumonia}.
\newblock Online; accessed 16 October 2020.

\bibitem{rubin2020others}
GD~Rubin, CJ~Ryerson, LB~Haramati, N~Sverzellati, and JP~Kanne.
\newblock others,“the role of chest imaging in patient management during the
  covid-19 pandemic: a multinational consensus statement from the fleischner
  society,”.
\newblock {\em Chest}, 2020.

\bibitem{bib:chen2020clinical}
Tao Chen, Di~Wu, Huilong Chen, Weiming Yan, Danlei Yang, Guang Chen, Ke~Ma,
  Dong Xu, Haijing Yu, Hongwu Wang, et~al.
\newblock Clinical characteristics of 113 deceased patients with coronavirus
  disease 2019: retrospective study.
\newblock {\em Bmj}, 368, 2020.

\bibitem{bib:borges2020novel}
Israel~J{\'u}nior Borges~do Nascimento, Nensi Cacic, Hebatullah~Mohamed
  Abdulazeem, Thilo~Caspar von Groote, Umesh Jayarajah, Ishanka Weerasekara,
  Meisam~Abdar Esfahani, Vinicius~Tassoni Civile, Ana Marusic, Ana Jeroncic,
  et~al.
\newblock Novel coronavirus infection ({COVID-19}) in humans: a scoping review
  and meta-analysis.
\newblock {\em Journal of clinical medicine}, 9(4):941, 2020.

\bibitem{bib:yang2020prevalence}
Jing Yang, Ya~Zheng, Xi~Gou, Ke~Pu, Zhaofeng Chen, Qinghong Guo, Rui Ji, Haojia
  Wang, Yuping Wang, and Yongning Zhou.
\newblock Prevalence of comorbidities and its effects in patients infected with
  {SARS-CoV-2}: a systematic review and meta-analysis.
\newblock {\em International Journal of Infectious Diseases}, 94:91--95, 2020.

\bibitem{bib:zhang2020d}
Litao Zhang, Xinsheng Yan, Qingkun Fan, Haiyan Liu, Xintian Liu, Zejin Liu, and
  Zhenlu Zhang.
\newblock D-dimer levels on admission to predict in-hospital mortality in
  patients with covid-19.
\newblock {\em Journal of Thrombosis and Haemostasis}, 18(6):1324--1329, 2020.

\bibitem{bib:petrilli2020factors}
Christopher~M Petrilli, Simon~A Jones, Jie Yang, Harish Rajagopalan, Luke
  O’Donnell, Yelena Chernyak, Katie~A Tobin, Robert~J Cerfolio, Fritz
  Francois, and Leora~I Horwitz.
\newblock Factors associated with hospital admission and critical illness among
  5279 people with coronavirus disease 2019 in new york city: prospective
  cohort study.
\newblock {\em bmj}, 369, 2020.

\bibitem{bib:henry2020hematologic}
Brandon~Michael Henry, Maria Helena~Santos De~Oliveira, Stefanie Benoit, Mario
  Plebani, and Giuseppe Lippi.
\newblock Hematologic, biochemical and immune biomarker abnormalities
  associated with severe illness and mortality in coronavirus disease 2019
  (covid-19): a meta-analysis.
\newblock {\em Clinical Chemistry and Laboratory Medicine (CCLM)},
  58(7):1021--1028, 2020.

\bibitem{bib:li2020elevated}
Chang Li, Jianfang Ye, Qijian Chen, Weihua Hu, Lingling Wang, Yameng Fan,
  Zhanjin Lu, Jie Chen, Zaishu Chen, Shiyan Chen, et~al.
\newblock Elevated lactate dehydrogenase ({LDH}) level as an independent risk
  factor for the severity and mortality of {COVID-19}.
\newblock {\em Aging (Albany NY)}, 12(15):15670, 2020.

\bibitem{bib:naymagon2020admission}
Leonard Naymagon, Nicole Zubizarreta, Jonathan Feld, Maaike van Gerwen,
  Mathilda Alsen, Santiago Thibaud, Alaina Kessler, Sangeetha Venugopal, Iman
  Makki, Qian Qin, et~al.
\newblock Admission {D-dimer} levels, {D-dimer} trends, and outcomes in
  {COVID-19}.
\newblock {\em Thrombosis Research}, 196:99--105, 2020.

\bibitem{bib:pradhan2020sex}
Ajay Pradhan and Per-Erik Olsson.
\newblock Sex differences in severity and mortality from {COVID-19}: are males
  more vulnerable?
\newblock {\em Biology of Sex Differences}, 11(1):1--11, 2020.

\bibitem{GradCAM}
Ramprasaath~R Selvaraju, Michael Cogswell, Abhishek Das, Ramakrishna Vedantam,
  Devi Parikh, and Dhruv Batra.
\newblock {Grad-CAM: Visual explanations from deep networks via gradient-based
  localization}.
\newblock In {\em Proceedings of the IEEE International Conference on Computer
  Vision}, pages 618--626, 2017.

\bibitem{latinne2001adjusting}
Patrice Latinne, Marco Saerens, and Christine Decaestecker.
\newblock Adjusting the outputs of a classifier to new a priori probabilities
  may significantly improve classification accuracy: evidence from a
  multi-class problem in remote sensing.
\newblock In {\em ICML}, volume~1, pages 298--305, 2001.

\bibitem{bib:ConceptDrift}
Jie Lu, Anjin Liu, Fan Dong, Feng Gu, Joao Gama, and Guangquan Zhang.
\newblock Learning under concept drift: {A} review.
\newblock {\em {IEEE Transactions on Knowledge and Data Engineering}},
  31(12):2346--2363, 2018.

\end{thebibliography}


\begin{thebibliography}{1}

\bibitem{kour2014real}
George Kour and Raid Saabne.
\newblock Real-time segmentation of on-line handwritten arabic script.
\newblock In {\em Frontiers in Handwriting Recognition (ICFHR), 2014 14th
  International Conference on}, pages 417--422. IEEE, 2014.

\bibitem{kour2014fast}
George Kour and Raid Saabne.
\newblock Fast classification of handwritten on-line arabic characters.
\newblock In {\em Soft Computing and Pattern Recognition (SoCPaR), 2014 6th
  International Conference of}, pages 312--318. IEEE, 2014.

\bibitem{hadash2018estimate}
Guy Hadash, Einat Kermany, Boaz Carmeli, Ofer Lavi, George Kour, and Alon
  Jacovi.
\newblock Estimate and replace: A novel approach to integrating deep neural
  networks with existing applications.
\newblock {\em arXiv preprint arXiv:1804.09028}, 2018.

\end{thebibliography}

\appendix
\section{First- and second-order features} \label{appendix}
We present in Tables \ref{tab:FirstOrder} and \ref{tab:SecondOrder} the formal definition of first- and second-order features introduced in section~\ref{subsec:handcrafted}.
The content of such tables adopt  the following notation:

\begin{itemize}
\item $X$ is a set of $N_p$ pixel in a ROI;
\item $S(i)$ is the first order histogram of the ROI using $N_g$ discrete intensity levels, equally spaced from 0 with a defined width of 0.1;
\item $ s(i) = \frac{S(i)}{N_p}$ is the normalized first order histogram;
\item $V_{pixel}$ is the volume of a pixel in mm;
\item $X_{10} $ is the $10^{th} $ percentile of X;
\item $X_{90} $ is the $90^{th} $ percentile of X;
\item $X_{10-90} $ is the image array with gray levels in between, or equal to the $10^{th}$ and  $90^{th} $ percentile of X;
\item $\overline X$ is the mean value of the image array;
\item $P(i,j)$ co-occurence matrix with a defined distance ($\delta$= 1) and angle ($\theta$=0);
\item $p(i,j) = \frac{P(i,j)}{\sum P(i,j)}$ is the normalized co-occurence matrix;
\item $p_x(i)= \sum_{j=1}^{N_g} P(i,j)$ and $p_y(i)= \sum_{i=1}^{N_g} P(i,j)$ are  the marginal probabilities per row and per column, respectively;
\item $\mu_x$ and $\mu_y$ are the mean grey level intensities, defined as Joined Average, of $p_x$ and $p_y$ respectively. If  $P(i,j)$  is symmetrical $p_x$ = $p_y$;
\item $\sigma_x$ and $\sigma_y$ are the standard deviations of $p_x$ and $p_y$ respectively;
\item $p_{x+y}(k) = \sum_{i=1}^{N_g} \sum_{j=1}^{N_g} p(i,j)$, where $i+j= k$, and $k=2,3,.., 2N_g$;
\item $p_{x-y}(k) = \sum_{i=1}^{N_g} \sum_{j=1}^{N_g} p(i,j)$, where $|i-j|= k$, and $k=0,1,.., N_g-1$;
\item HX, HY and HXY are the entropy of $ p_x$, $ p_y$ and $ p(i,j)$, respectively.
\item $ HXY1= -\sum_{i=1}^{N_g}\sum_{j=1}^{N_g}p(i,j) \cdot \log[p_x(i)p_y(j)] $  is an auxiliary quantity;
\item $ HXY2= -\sum_{i=1}^{N_g}\sum_{j=1}^{N_g}p_x(i)p_y(j) \cdot \log[p_x(i)p_y(j)] $ is an auxiliary quantity;
\item DA is the Difference Average used to obtain the Difference Variance;

\end{itemize}

\renewcommand{\thetable}{A.\arabic{table}}

\setcounter{table}{0}
\begin{table*}[]
\centering
\normalsize
\resizebox{0.7\textwidth}{!}{
\begin{tabular}[t]{l|l}
\toprule
\bf{Feature} & \bf{Definition} \\
\midrule
\midrule
Energy & $ \sum_{i=0}^{N_p} X(i)^2$ \\
\midrule
Total Energy & $ V_{pixel} \cdot \sum_{i=0}^{N_p} X(i)^2$  \\
\midrule
Entropy & $ - \sum_{i=1}^{N_g} s(i) \cdot \log[s(i)]$, for $s(i) >0 $ \\
\midrule
Minimum & $ min(X)$ \\
\midrule
Maximum  & $ max(X)$ \\
\midrule
Mean & $ \frac{1}{N_p}\sum_{i=1}^{N_p} X(i)$ \\
\midrule
Median & median grey level intensity  \\
\midrule
Interquartile Range  &  $X_{75}- X_{25}  $ \\
\midrule
Range &  $ max(X) - min(X) $\\
\midrule
Mean Absolute Deviation & $ \frac{1}{N_p} \cdot \sum_{i=1}^{N_p} |X(i)- \overline X |  $\\
\midrule
Robust Mean Absolute Deviation & $ \frac{1}{N_{10-90}} \cdot \sum_{i=1}^{N_{10-90}} |X_{10-90}(i)- \overline X_{10-90} |  $\\
\midrule
Root Mean Squared &  $ \sqrt{(\frac{1}{N_p} \cdot \sum_{i=1}^{N_p} X(i)^2)} $\\
\midrule
Skewness & $ \frac{\frac{1}{N_p} \cdot \sum_{i=1}^{N_p} (X(i) -\overline X )^3}{(\sqrt{\frac{1}{N_p} \cdot \sum_{i=1}^{N_p} (X(i) -\overline X )^2})^3}$ \\
\midrule
Kurtosis & $ \frac{\frac{1}{N_p} \cdot \sum_{i=1}^{N_p} (X(i) -\overline X )^4}{(\frac{1}{N_p} \cdot \sum_{i=1}^{N_p} (X(i) -\overline X )^2)^2}$\\
\midrule
Variance & $ \frac{1}{N_p} \cdot \sum_{i=1}^{N_p} (X(i) -\overline X )^2 $ \\
\midrule
Uniformity & $ \sum_{i=1}^{N_g} s(i)^2 $ \\
\bottomrule
\end{tabular}}
\caption{Definition of the first-order statistical measures.}
\label{tab:FirstOrder}

\end{table*}

\begin{table*}[]
\normalsize
\centering
\resizebox{.99\textwidth}{!}{

\begin{tabular}[t]{l|l}

\toprule
\bf{Feature} & \bf{Definition} \\
\midrule
\midrule
Sum Squares & $\sum_{i=1}^{N_g} \sum_{j=1}^{N_g} (i- \mu_x )^2 \cdot p(i,j)$\\ 
\midrule
Sum Entropy & $ \sum_{k=2}^{2N_g} p_{x+y}(k) \cdot \log[p_{x+y}(k)]$, for $p_{x+y}(k) > 0$\\ 
\midrule
Sum Average & $ \sum_{k=2}^{2N_g} p_{x+y}(k)k$\\ 
\midrule
MCC (Maximal Correlation Coefficient) & $ \sqrt{\textrm{second largest eigenvalue of Q}}$, where $ Q(i,j)= \sum_{k=0}^{N_g} \frac{p(i,k)p(j,k)}{p_x(i)p_y(k)}$ \\ 
\midrule
Maximum Probability & $ \max(p(i,j))$\\ 
\midrule
Joint Entropy & $\sum_{i=1}^{N_g} \sum_{j=1}^{N_g} p(i,j) \log[p(i,j)]$ , for $p(i,j) > 0 $\\ 
\midrule
Joint Energy & $\sum_{i=1}^{N_g} \sum_{j=1}^{N_g} (p(i,j))^2$\\ 
\midrule
Joint Average & $\sum_{i=1}^{N_g} \sum_{j=1}^{N_g} p(i,j)i$\\ 
\midrule
Inverse Variance & $ \sum_{k=0}^{N_g-1} \frac{p_{x-y}(k)}{k^2}$\\ 
\midrule
IMC (Informational Measure of Correlation) 2  & $ \sqrt{1- e^{-2(HXY2- HXY)}}$\\ 
\midrule
IMC (Informational Measure of Correlation) 1 & $ \frac{HXY-HXY1}{max\{HX,HY\}}$\\ 
\midrule
IDN (Inverse Difference Normalized) & $ \sum_{k=0}^{N_g-1} \frac{p_{x-y}(k)}{1+(\frac{k}{N_g})}$\\ 
\midrule
IDM (Inverse Difference Moment) & $ \sum_{k=0}^{N_g-1} \frac{p_{x-y}(k)}{1+k^2}$\\ 
\midrule
ID (Inverse Difference) & $ \sum_{k=0}^{N_g-1} \frac{p_{x-y}(k)}{1+k}$\\ 
\midrule
Difference Variance & $ \sum_{k=0}^{N_g-1} (k- DA)^2 \cdot p_{x-y}(k)$\\ 
\midrule
Difference Entropy & $ \sum_{k=0}^{N_g-1} k \cdot p_{x-y}(k) \log[p_{x-y}(k)]$ , for $p_{x-y}(k) > 0 $\\ 
\midrule
Difference Average & $ \sum_{k=0}^{N_g-1} k \cdot p_{x-y}(k)$ \\ 
\midrule
Correlation & $\frac{\sum_{i=1}^{N_g} \sum_{j=1}^{N_g} p(i,j) \cdot i \cdot j - \mu_x \mu_y}{\sigma_x(i) \sigma_y(j)}$  \\ 
\midrule
Contrast & $\sum_{i=1}^{N_g} \sum_{j=1}^{N_g} (i-j)^2 \cdot p(i,j)$ \\ 
\midrule
Cluster Tendency & $\sum_{i=1}^{N_g} \sum_{j=1}^{N_g} (i+j- \mu_x - \mu_y)^2 \cdot p(i,j)$ \\ 
\midrule
Cluster Shade & $\sum_{i=1}^{N_g} \sum_{j=1}^{N_g} (i+j- \mu_x - \mu_y)^3 \cdot p(i,j)$ \\ 
\midrule
Cluster Prominence & $\sum_{i=1}^{N_g} \sum_{j=1}^{N_g} (i+j- \mu_x - \mu_y)^4 \cdot p(i,j)$\\ 

\bottomrule

\end{tabular}}

\caption{Definition of the second-order statistical measures.}
\label{tab:SecondOrder}
\end{table*}

\end{document}